\documentclass[11pt]{iopart}

\expandafter\let\csname equation*\endcsname\relax
\expandafter\let\csname endequation*\endcsname\relax
\usepackage{amsmath}

\usepackage{graphicx}
\usepackage{rotating}
\usepackage{cite}
\usepackage{color}
\usepackage{soul}
\usepackage{siunitx}
\usepackage[colorlinks=true
,urlcolor=blue
,anchorcolor=blue
,citecolor=blue
,filecolor=blue
,linkcolor=blue
,menucolor=blue
,pagecolor=blue
,linktocpage=true
,pdfproducer=medialab
,pdfa=true
]{hyperref}
\usepackage{soul}
\usepackage[table]{xcolor}
\usepackage{multirow}

\begin{document}
	
	\setlength{\parindent}{0pt}
	
	\title[ ]{Electron power absorption in capacitively coupled neon-oxygen plasmas: a comparison of experimental and computational results
	}
	
	\author{ A. Derzsi$^{1}$, P. Hartmann$^1$, M. Vass$^{1,2}$, B. Horv\'ath$^1$, M. Gyulai$^1$, I. Korolov$^2$, J. Schulze$^{2,3}$,  Z. Donk\'o$^1$}
	
	\address{
		$^1$ Institute for Solid State Physics and Optics, Wigner Research Centre for Physics, 1121 Budapest, Hungary\\
		$^2$ Department of Electrical Engineering and Information Science, Ruhr University Bochum, 44780 Bochum, Germany\\
		$^3$ Key Laboratory of Materials Modification by Laser, Ion and Electron Beams, School of Physics, Dalian University of Technology, 116024 Dalian, China
	}
	
	\ead{derzsi.aranka@wigner.hu}
	
	\begin{abstract}
		Phase Resolved Optical Emission Spectroscopy (PROES) measurements combined with 1d3v Particle-in-Cell/Monte Carlo Collisions (PIC/MCC) simulations are used to study the electron power absorption and excitation/ionization dynamics in capacitively coupled plasmas (CCPs) in mixtures of neon and oxygen gases. The study is performed for a geometrically symmetric CCP reactor with a gap length of 2.5~cm at a driving frequency of 10~MHz and a peak-to-peak voltage of 350~V. The pressure of the gas mixture is varied between 15~Pa and 500~Pa, while the neon/oxygen concentration is tuned between 10\% and 90\%. For all discharge conditions, the spatio-temporal distribution of the electron-impact excitation rate from the Ne ground state into the Ne~$\rm{2p^53p_0}$ state measured by PROES and obtained from PIC/MCC simulations show good qualitative agreement. Based on the emission/excitation patterns, multiple operation regimes are identified. Localized bright emission features at the bulk boundaries, caused by local maxima in the electronegativity are found at high pressures and high O$_2$ concentrations. The relative contributions of the ambipolar and the Ohmic electron power absorption are found to vary strongly with the discharge parameters: the Ohmic power absorption is enhanced by both the high collisionality at high pressures and the high electronegativity at low pressures.
		In the wide parameter regime covered in this study, the PROES measurements are found to accurately represent the ionization dynamics, i.e., the discharge operation mode. This work represents also a successful experimental validation of the discharge model developed for neon-oxygen CCPs.

	\end{abstract}
	
\section{Introduction} \label{sec:Introduction}

Capacitively Coupled Plasmas (CCPs) represent important reactor types widely used in plasma etching, sputtering, deposition, and cleaning processes providing the  basis for semiconductor manufacturing and biomedical applications \cite{Liebermann_book,Makabe_book,Chabert_book,Makabe08,Makabe_2019}. In such plasmas, efficient control of the particle properties at the electrodes (e.g., the fluxes of chemically reactive species and the ion flux-energy distribution), which determine the plasma--surface interaction, is a basic requirement. In order to achieve improved process control in such systems, a detailed understanding of the electron power absorption and ionization dynamics is necessary. 

The spatio-temporal distributions of the electron power absorption and the ionization within the radiofrequency (RF) period determine the operation mode of the discharge. In low-pressure CCPs, several different discharge operation modes can be identified: the $\alpha$-mode and the $\gamma$-mode \cite{Belenguer1990} are typical in electropositive gases, while the drift-ambipolar (DA) mode \cite{Schulze2011} and the striation (STR) mode \cite{Liu2016, Liu2017, 
Wang_2019, Skarphedinsson_2020, Proto_2021} can be observed in electronegative gases. In the $\alpha$-mode, the ionization, caused by energetic electrons accelerated by electric fields during the times of sheath expansion, is concentrated at the bulk side of the expanding sheath edge, while in the $\gamma$-mode, the ionization, dominated by secondary electrons (SEs) accelerated by the strong electric field inside the sheaths, is concentrated within the sheath region. In the DA-mode, the ionization can be observed across the whole bulk region and at the collapsing sheath edges, as it is generated by fast electrons accelerated by the strong drift electric field in the plasma bulk caused by the low conductivity of the plasma bulk, and by the ambipolar electric fields at the sheath edges caused by the strong gradients of the electron density. In the STR-mode, which develops when both positive and negative ions can react to the fast variation of the RF electric field, the ionization, concentrated within the bulk region, exhibits layered structures called ``striations''; these are generated as a result of the formation of alternating space charges and the modulation of the electric field and the energy gain of electrons in the plasma bulk. The simultaneous presence of the ionization patterns characteristic of the different discharge operation modes can be observed in low-pressure CCPs under specific discharge conditions, as well as transitions between these modes by varying the external control parameters \cite{Derzsi_2017,Gudmundsson_2019,Donko_2017_PPCF,Brandt_2019,Hyo-Chang_2019_O2,Gibson_2017}. 

Experimental observation of the space and time-resolved plasma emission based on Phase Resolved Optical Emission Spectroscopy (PROES) \cite{Gans_2004, Gans_2010, Schulze_JPD_2010} provides invaluable information on the electron power absorption and excitation/ionization dynamics in low-pressure CCPs \cite{Schulze2011,Liu2016}. For PROES measurements, Ne is often used as a tracer gas due to its favorable spectroscopic properties. By adding Ne in a small concentration (typically 5\% -- 15\%) to the background gas, and measuring the emission from a carefully selected atomic transition (e.g., Ne~$\rm{2p^5(^2P^o_{1/2})3p_0} \rightarrow \rm{2p^5(^2P^o_{1/2})3s_1}$ with a wavelength of 585.25 nm) the spatio-temporal electron impact excitation rate from the ground state into the upper level can be derived. This way, by selecting an emission line resulting from an excited state with a high electron impact excitation threshold energy, information about the dynamics of high-energy electrons (which are typically responsible both for the excitation and the ionization processes in the discharge) can be obtained, making PROES an effective non-intrusive diagnostic technique. Although PROES provides information about the spatio-temporal distribution of the electron-impact excitation dynamics from the ground state into the selected excited atomic state in the discharge, it is generally considered to probe the discharge operation mode (which, in turn, is determined by the spatio-temporal distribution of the ionization dynamics) as well.

Recently, the applicability of PROES to probe the discharge operation mode was investigated in low-pressure CCPs operated in pure neon \cite{Horvath_2020}. In this work, a detailed comparison of computational and experimental results focusing on the spatio-temporal distributions of the excitation and the ionization rates in geometrically symmetric single-frequency neon CCPs has been performed in a wide parameter regime. Particle-in-Cell/Monte Carlo Collisions (PIC/MCC) simulations \cite{Hockney_Book, Birdsall_Book, Birdsall_1991, Schweigert_1999, Diomede_2005, Verboncoeur2005, Schneider, Radmilovic-Radjenovic2009, Donko_2011_PSST, Donko2012, Gudmundsson_2013, Sun_2016} and PROES measurements were carried out at driving frequencies ranging from 3.39~MHz to 13.56~MHz and pressures from 60~Pa to 500~Pa, at a fixed peak-to-peak voltage of 330~V and an electrode gap of 2.5~cm. At fixed frequencies, transitions of the discharge operation mode from the $\alpha$-mode to the $\gamma$-mode were observed by increasing the pressure. The electron impact excitation rates from the ground state into the Ne~$\rm{2p^5(^2P^o_{1/2})3p_0}$ state (for which we will use the simplified notation Ne~$\rm{3p_0}$) obtained from PROES measurements and PIC/MCC simulations were in good agreement in all cases. 
However, it was found that PROES results do not always probe the ionization and significant $\gamma$-mode ionization can take place in the discharge even in the cases when this is not seen in the spatio-temporal distribution of the Ne~$\rm{3p_0}$ excitation. This was explained by the difference in the cross sections of the electron impact excitation into the observed level and the ionization as a function of the electron energy \cite{Horvath_2020}. Although the threshold energy of the Ne~$\rm{3p_0}$ excitation process is close to the one of the ionization, the shapes of the cross sections are different within the electron energy range of the discharges studied. As a consequence of this, it is more likely that the highly energetic SEs within the sheaths induce ionization rather than excitation.

Due to their high relevance in material processing, oxygen CCPs have been studied extensively, both experimentally and by simulations \cite{Bronold_2007, Dittmann_2007, Matyash_2007, Bera_2011,Teichmann_2013, Liu_2013,Liu_2015,Benyoucef_2015, Kullig_2015, Hannesdottir_2016, Hannesdottir_2017, Gudmundsson_2017, Gudmundsson_2018,Derzsi_2016,Derzsi_2017,Wang_2020, Vass_2020}. In pure oxygen CCPs, transitions between the $\alpha$-mode and the DA-mode have been found by changing the gas pressure \cite{ Derzsi_2017,Gudmundsson_2019}, the gap distance \cite{Gudmundsson_2019, Hyo-Chang_2019_O2}, the driving frequency \cite{Derzsi_2017,Gudmundsson_2018}, the driving voltage waveform \cite{Derzsi_2015, Derzsi_2017, Gibson_2017, Gudmundsson_2017, Donko_2017_PPCF, Donko_2018}, and the external magnetic field \cite{Wang_2020}.
A transition from a hybrid DA-$\alpha$ mode to a pure $\alpha$-mode was observed by increasing the driving frequency at a constant pressure \cite{Gudmundsson_2018}, as well as by increasing the pressure or the electrode gap \cite{Gudmundsson_2019}. PIC/MCC simulations and PROES measurements showed a transition from the DA-mode to the $\alpha$-mode in oxygen CCPs driven by tailored voltage waveforms by changing the number of consecutive harmonics included in the driving voltage waveform at high base frequencies \cite{Derzsi_2017}.

In this work, we study the electron power absorption and excitation/ionization dynamics in capacitively coupled plasmas (CCPs) operated in mixtures of neon and oxygen. PROES measurements are combined with PIC/MCC simulations in a wide parameter regime. The study is performed in a geometrically symmetric CCP reactor with a gap length of 2.5~cm, at a driving frequency of 10~MHz and a peak-to-peak voltage of 350~V. The pressure of the gas mixture is varied between 15~Pa and 500~Pa, while the mixing ratio of  neon-to-oxygen is between 10\% and 90\%. Transitions between different discharge operation modes are found to be induced by changing the gas pressure and varying the neon/oxygen concentration ratio in the discharge, as well as the development of localized bright emission/excitation features at the edges of the bulk region at high pressures and high O$_2$ concentrations.
The good qualitative agreement between the PROES results and the PIC/MCC simulation results supports the validity of the discharge model developed for neon-oxygen CCPs. While experimental validation of simulations is of major importance for the reliability of the simulation results, this has only been done for some inert gases \cite{Horvath_2020, Schulenberg21} and electronegative gases \cite{Derzsi_2016}, but such efforts are barely found for reactive gas mixtures \cite{Donko_2017_PPCF}, which are required for process development.

The paper is structured in the following way. In section~\ref{sec:Experimental}, the experimental setup is presented. In section~\ref{sec:Simulation}, the simulation method and the neon-oxygen discharge model is described, including details on the collision processes in section~\ref{sec:Collisions}, the surface processes in section~\ref{sec:SurfaceProcesses}, the gas heating module in section~\ref{sec:Heating}, the calculation of the density of $\rm{O_2 (a^1 \Delta_g)}$ metastable molecules in section~\ref{sec:Metastable}, the study of the electron power absorption based on the Boltzmann term method in section~\ref{sec:EPowerAbsorption}, and the simulation settings in section~\ref{sec:Sim_settings}. The results are shown and discussed in section~\ref{sec:Results}. The conclusions are drawn in section~\ref{sec:Conclusions}.

\section{Experimental setup and discharge conditions} \label{sec:Experimental}

The space and time resolved optical emission of the discharge is measured experimentally, by PROES \cite{Gans_2004, Gans_2010, Schulze_JPD_2010}, in which the emitted light from the selected excited atomic/molecular state is spatio-temporally resolved. From that, the electron impact excitation rate from the ground state into the observed level, $W_{0,i} (x, t)$, can be calculated, as introduced in \cite{Schulze_JPD_2010}. In order to perform a successful PROES measurement on a CCP, some crucial conditions have to be fulfilled \cite{Schulze_JPD_2010}:
(i) Knowledge of the optical transition rates in the gas is needed.
(ii) The contribution of cascades, excitation from metastable levels and quenching to the population of the measured excited state need to be negligibly low.
(iii) The intensity of the emitted light at the measured line has to be high enough. 
(iv) No superposition with other optical lines is tolerated within the spectral resolution of the spectrometer or the interference filter.
(v) The lifetime of the relevant excited state has to be short enough to temporally resolve the RF period (100~ns in the current study).
All in all, the choice of the optical line to measure is critical. A typical line that satisfies these criteria well and is, therefore, often applied for PROES is the emission line corresponding to the Ne~$\rm{2p^5(^2P^o_{1/2})3p_0} \rightarrow \rm{2p^5(^2P^o_{1/2})3s_1}$ transition, with a wavelength of 585.25~nm and lifetime of 16.26 ns~\cite{T6389}. Its threshold energy for electron impact excitation from the ground state is 18.966~eV~\cite{Saloman04}.

\begin{figure}
	\centering
	\includegraphics[width=\linewidth]{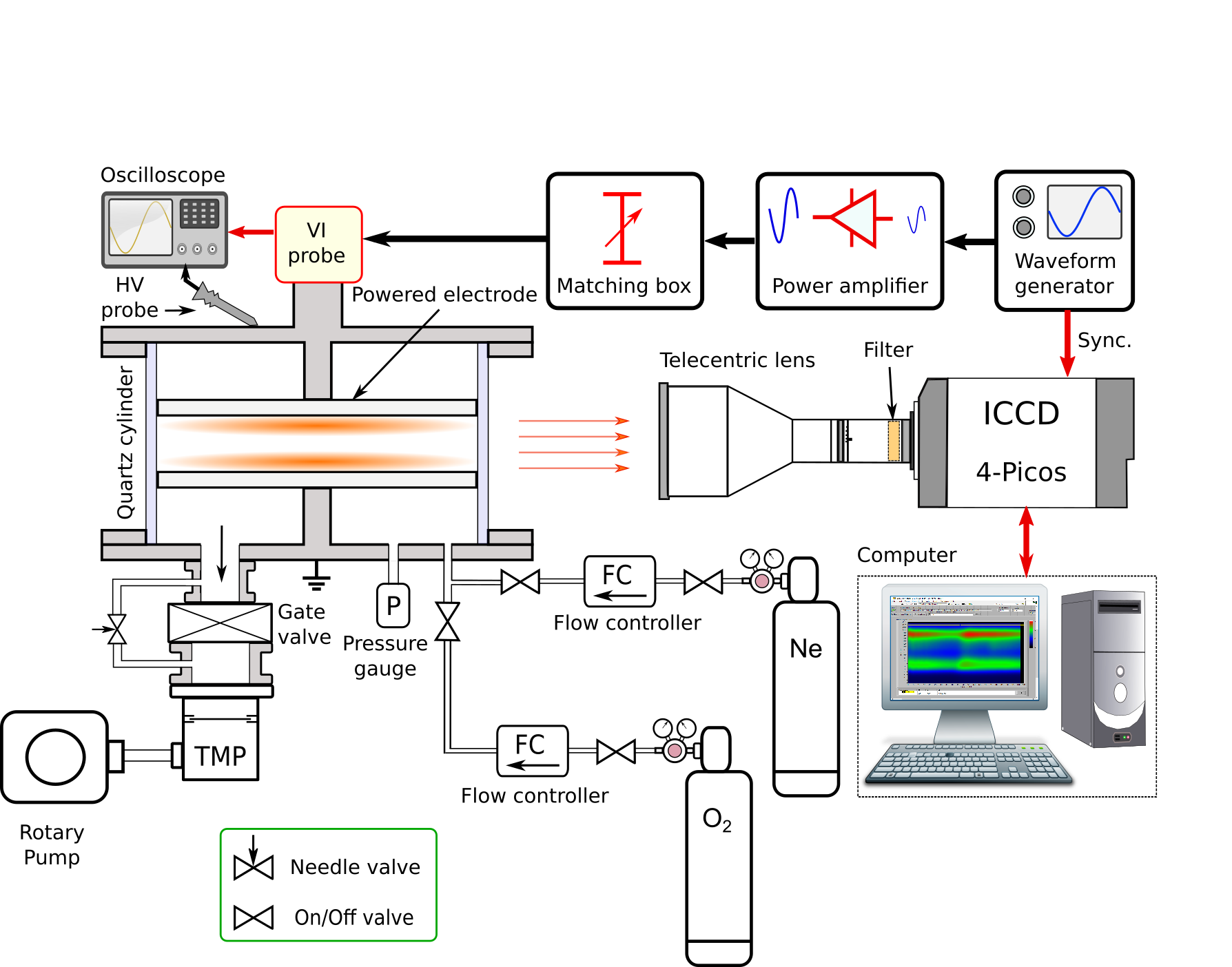}
	\caption{Scheme of the experimental setup.}
	\label{fig:expbudneonv2}
\end{figure}

For the current measurements, our geometrically symmetric ``Budapest v.3'' plasma reactor is used, which was first introduced in \cite{Horvath_2020}. A sketch of the setup is shown in figure~\ref{fig:expbudneonv2}. The flat disk electrodes made of 304 grade stainless steel with identical diameters of 14.0 cm are situated within a quartz cylinder. Their separation is 2.5 cm. The upper electrode is driven by RF voltage, while the lower one is grounded. There is no active cooling applied on the electrodes.
The chamber can be evacuated via a turbomolecular and a rotary pump, providing a base pressure of approximately $10^{-5}$~Pa. The experiments are performed in a gas flow of $\approx$3~sccm set by two flow controllers.
The driving voltage is provided by a waveform generator (Juntek JDS-2900), which is connected to a linear power amplifier (RM BLA-300) and an impedance matching box (MFJ-949E). The voltage drop across the discharge is measured by a high voltage probe (HP 10076A, 100:1) right at the power feeding point, and the pressure is monitored by capacitive gauges (Pfeiffer Vacuum CMR264 and MKS Baratron
631A).

PROES measurements are performed with a fast-gateable ICCD camera (4 Picos, Stanford Computer Optics). The optical emission from the Ne~$\rm{3p_0}$ state with a wavelength of 585.25~nm is measured. In order to limit the measured light to this specific line, an interference filter is applied with a central wavelength of 585~nm and a spectral full width at half maximum of $\sim$10~nm. The gate width of the camera is set to 2~ns. The camera is equipped with a Thorlabs MVTC23013 0.128x bi-telecentric lens, by which two-dimensional pictures can be taken. The spatial resolution is approximately 150 $\rm{\mu m}$. Due to the lateral uniformity of the plasma, the data are averaged in the direction perpendicular to the discharge axis, which reduces the noise significantly.

\section{Simulation method} \label{sec:Simulation}

The simulations are based on a one dimensional in space and three dimensional in velocity space (1d3v) Particle-in-Cell/Monte Carlo Collisions (PIC/MCC) simulation code  \cite{Hockney_Book, Birdsall_Book, Birdsall_1991, Schweigert_1999, Diomede_2005, Verboncoeur2005, Schneider, Radmilovic-Radjenovic2009, Donko_2011_PSST, Donko2012, Gudmundsson_2013, Sun_2016}. This code (named \emph{PICit!}), recently developed in our group, is suitable to model geometrically symmetric capacitively coupled RF discharges in various mixtures of electropositive/electronegative gases. The particles traced in the simulations of neon-oxygen gas mixture plasmas are electrons, Ne$^+$ ions, fast Ne atoms (Ne$^{\rm f}$), ${\rm O_2^+}$ ions, ${\rm O^-}$ ions and fast ${\rm O_2}$ molecules (O$_2^{\rm f}$). The set of collision processes taken into account in the neon-oxygen discharge model is based on the sets of collision processes used previously for simulation studies of CCP discharges operated in pure neon \cite{Horvath_2020} and oxygen \cite{Derzsi_2015,Derzsi_2017,Donko_2018,Hyo-Chang_2019_O2,Vass_2020, Vass_2021}, complemented with ``cross processes'' between oxygen and neon species and collision processes for fast neutrals (see section~\ref{sec:Collisions}).
The heating of the background gas due to elastic collisions of fast neutrals and ions with thermal (background) atoms/molecules, as well as heating up of the electrodes due to inelastic collisions of plasma particles with the electrodes \cite{Geonwoo_heating} is taken into account in the model (see section~\ref{sec:Heating}). The complexity of the model necessitates a relatively detailed description, which we provide below.

\subsection{Collision processes}\label{sec:Collisions}

The elementary collision processes included in the discharge model are listed in table~\ref{table:Neon-Oxygen-mixture}.

\begin{table}[p]
\caption{\label{n2} List of collision processes for neon--oxygen mixtures.
}
\footnotesize
\begin{tabular}{@{}llll}
\br
\#&Reaction&Process&References\\
\mr
1& $ \rm{e^{-} + Ne \longrightarrow e^{-} + Ne} $  &   Elastic scattering & \cite{Biagi7.1}\\

2& $ \rm{e^{-} + Ne \longrightarrow e^{-} + Ne} $  &   $\rm{2p^5(^2P^o_{3/2})3s_2}$ excitation & \cite{Biagi7.1}\\

3& $ \rm{e^{-} + Ne \longrightarrow e^{-} + Ne} $  &   $\rm{2p^5(^2P^o_{3/2})3s_1}$ excitation & \cite{Biagi7.1}\\

4&$ \rm{e^{-} + Ne \longrightarrow e^{-} + Ne} $  &   $\rm{2p^5(^2P^o_{1/2})3s_0}$ excitation & \cite{Biagi7.1}\\

5&$ \rm{e^{-} + Ne \longrightarrow e^{-} + Ne} $  &   $\rm{2p^5(^2P^o_{1/2})3s_1}$ excitation & \cite{Biagi7.1}\\

6&$ \rm{e^{-} + Ne \longrightarrow e^{-} + Ne} $  &   $\rm{\sum 2p^53p}$ excitation & \cite{Biagi7.1}\\

7&$ \rm{e^{-} + Ne \longrightarrow e^{-} + Ne} $  &   $\rm{\sum 2p^54s}$ excitation & \cite{Biagi7.1}\\

8&$ \rm{e^{-} + Ne \longrightarrow e^{-} + Ne} $  &   $\rm{\sum 2p^53d}$ excitation & \cite{Biagi7.1}\\

9&$ \rm{e^{-} + Ne \longrightarrow e^{-} + Ne} $  &   $\rm{\sum 2p^54p}$ excitation & \cite{Biagi7.1}\\

10&$ \rm{e^{-} + Ne \longrightarrow e^{-} + Ne} $  &   $\rm{2p^5(^2P^o_{1/2})3p_0}$ excitation & \cite{Biagi7.1}\\

11&$ \rm{e^{-} + Ne \longrightarrow 2 e^{-} + Ne^+} $  &   Ionization & \cite{Biagi7.1}\\
12& $ \rm{e^{-}+O_{2} \longrightarrow O_{2}+e^{-}} $  &   Elastic scattering & \cite{Biagi8.9}\\
13& $ \rm{e^{-}+O_{2}} (\textit{r}=0) \longrightarrow \rm{e^{-} + O_{2}} (\textit{r}>0) $  &   Rotational excitation & \cite{vahedi1995monte}\\
14& $ \rm{e^{-}+O_{2}} (\textit{v}=0) \longrightarrow \rm{e^{-} + O_{2}} (\textit{v}=1) $  &   Vibrational excitation & \cite{vahedi1995monte}\\
15&$ \rm{e^{-}+O_{2}} (\textit{v}=0) \longrightarrow \rm{e^{-} + O_{2}} (\textit{v}=2) $  &   Vibrational excitation & \cite{vahedi1995monte}\\
16&$ \rm{e^{-}+O_{2}} (\textit{v}=0) \longrightarrow \rm{e^{-} + O_{2}} (\textit{v}=3) $  &   Vibrational excitation & \cite{vahedi1995monte}\\
17&$ \rm{e^{-}+O_{2}} (\textit{v}=0) \longrightarrow \rm{e^{-} + O_{2}} (\textit{v}=4) $  &   Vibrational excitation & \cite{vahedi1995monte}\\
18&$ \rm{e^{-}+O_{2}}  \longrightarrow e^{-} + O_{2}   (a^{1}\Delta_{g}) $  &   Metastable excitation  (0.98 eV) & \cite{vahedi1995monte}\\
19&$ \rm{e^{-}+O_{2}}  \longrightarrow e^{-} + O_{2}   (b^{1}\Sigma_{g}) $  &   Metastable excitation  (1.63 eV) & \cite{vahedi1995monte}\\
20&$ \rm{e^{-}+O_{2}}  \longrightarrow  O + O^{-}  $  &  Dissociative attachment & \cite{vahedi1995monte}\\
21&$ \rm{e^{-}+O_{2}}  \longrightarrow  e^{-} + O_{2}  $  &  Excitation (4.5 eV) & \cite{vahedi1995monte}\\
22&$ \rm{e^{-}+O_{2}}  \longrightarrow  O(^{3}P) + O(^{3}P) + e^{-}  $  &  Dissociation (6.0 eV) & \cite{vahedi1995monte}\\
23&$ \rm{e^{-}+O_{2}}  \longrightarrow  O(^{3}P) + O(^{1}D) + e^{-}  $  &  Dissociation (8.4 eV) & \cite{vahedi1995monte}\\
24&$ \rm{e^{-}+O_{2}}  \longrightarrow  O(^{1}D) + O(^{1}D) + e^{-}  $  &  Dissociation (9.97 eV) & \cite{vahedi1995monte}\\
25&$ \rm{e^{-}+O_{2}}  \longrightarrow  O^{+}_{2} + e^{-} + e^{-} $  &  Ionization & \cite{gudmundsson2013benchmark}\\
26&$ \rm{e^{-}+O_{2}}  \longrightarrow  e^{-} + O + O(3p ^{3}P) $  &  Dissociative excitation (14.7 eV) & \cite{vahedi1995monte}\\
27&$ \rm{e^{-}+O^{-}}  \longrightarrow  e^{-} + e^{-} + O $  &  Electron impact detachment & \cite{gudmundsson2013benchmark}\\
28&$ \rm{e^{-}+O^{+}_{2}}  \longrightarrow  O(^{3}P) + O(^{1}D) $  &  Dissociative recombination & \cite{gudmundsson2013benchmark}\\

\mr

29&$ \rm{Ne^{+} + Ne \longrightarrow Ne^{+} + Ne} $  &   Isotropic scattering & \cite{PhepsNeonJILA} \\

30&$ \rm{Ne^{+} + Ne \longrightarrow Ne^{+} + Ne} $  &   Backscattering & \cite{PhepsNeonJILA} \\

31& Ne$^+$ + O$_2$ $\longrightarrow$ Ne + O$_2^+$ &  Charge transfer  & \cite{adams1972thermal}\\
32& Ne$^+$ + O$_2$ $\longrightarrow$ Ne$^+$ + O$_2$ &  Isotropic elastic scattering  & \cite{Langevin}\\

33&$ \rm{O^{+}_{2}+O_{2}}  \longrightarrow  O_{2}+ O^{+}_{2} $  &  Elastic scattering: charge exchange & \cite{gudmundsson2013benchmark}\\
34&$ \rm{O^{+}_{2}+O_{2}}  \longrightarrow  O_{2}+ O^{+}_{2} $  &  Elastic scattering: isotropic part & \cite{gudmundsson2013benchmark}\\

35& O$_2^+$ + Ne $\longrightarrow$ O$_2^+$ + Ne &  Isotropic elastic scattering  & \cite{Langevin}\\

36&$ \rm{O^{-}+O_{2}}  \longrightarrow  O^{-}+ O_{2} $  &  Elastic scattering & \cite{gudmundsson2013benchmark}\\
37&$ \rm{O^{-}+O_{2}}  \longrightarrow  O+ O_{2}+e^{-} $  &  Detachment & \cite{gudmundsson2013benchmark}\\
38&$ \rm{O^{-}+O_{2}^{+}}  \longrightarrow  O+ O_{2} $  &  Mutual neutralization & \cite{gudmundsson2013benchmark}\\
39&$ \rm{O^{-}+O_{2}(a^{1}\Delta_{g})}  \longrightarrow  O_{3}+e^{-} $  &  Associative detachment & \cite{bronold2007radio}\\

40& O$^-$ + Ne $\longrightarrow$ O$^-$ + Ne &  Isotropic elastic scattering  & \cite{Langevin}\\
41& O$^-$ + Ne$^+$ $\longrightarrow$ O + Ne &  Mutual neutralisation  & \cite{gudmundsson2013benchmark}\\

\mr

42&$ \rm{Ne^{f} + Ne \longrightarrow Ne^{f} + Ne} $  &   Elastic scattering &  \\

43&$ \rm{O^{f}_{2}+O_{2}}  \longrightarrow  O^{f}_{2}+O_{2} $  &  Elastic scattering & \\

44&$ \rm{Ne^{f} + O_{2} \longrightarrow Ne^{f} + O_{2}} $  &   Elastic scattering & \\

45&$ \rm{O^{f}_{2}+Ne  \longrightarrow  O^{f}_{2}+Ne }$  &  Elastic scattering & \\

\br
\end{tabular}\\
\label{table:Neon-Oxygen-mixture}
\end{table}

\begin{figure}[ht]
	\centering
	\includegraphics[width=.49\linewidth]{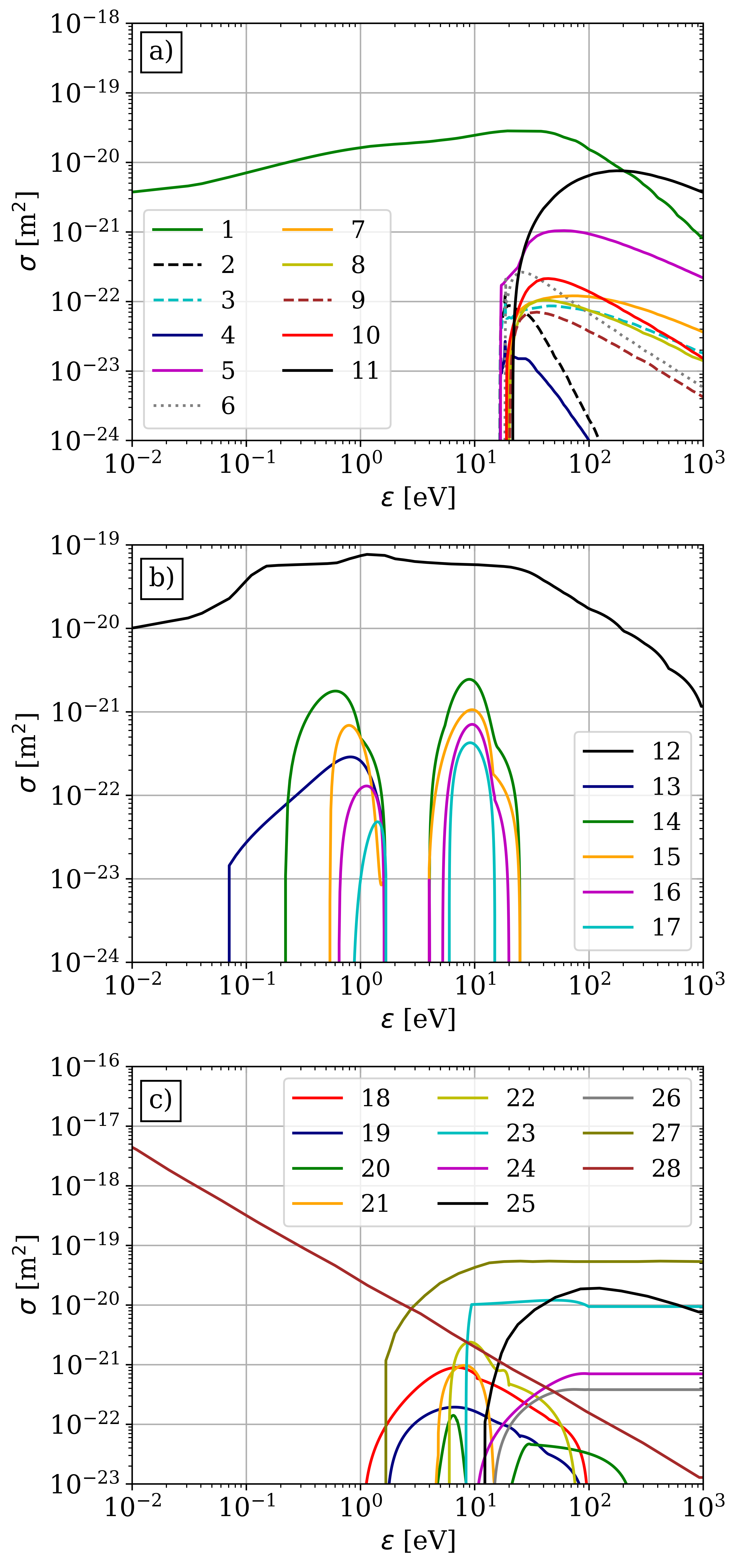}
	\caption{Cross sections of electron collisions listed in table~\ref{table:Neon-Oxygen-mixture} (processes 1--28) as a function of the kinetic energy of the projectile: 
	(a) Elastic scattering between electrons and Ne atoms (processes 1), electron impact excitation (processes 2-10) and ionization (process 11) of Ne. The red curve (process 10) in panel (a) corresponds to the excitation from the ground state into the Ne~$\rm{3p_0}$ state. 
	(b) Elastic scattering between electrons and O$_2$ molecules (processes 12), electron impact rotational excitation (process 13) and vibrational excitation (processes 14--17) of O$_2$.
	(c) Metastable excitation (processes 18 and 19), dissociative attachement (process 20), electronic excitation (process 21), dissociation (processes 22--24), ionization (process 25) and dissociative excitation (process 26) between electrons and O$_2$ molecules, as well as detachment with O$^-$ (process 27) and dissociative recombination with O$_2^+$ (process 28).}
	\label{fig:electron_collisions}
\end{figure}

\begin{figure}[ht]
	\centering
	\includegraphics[width=.5\linewidth]{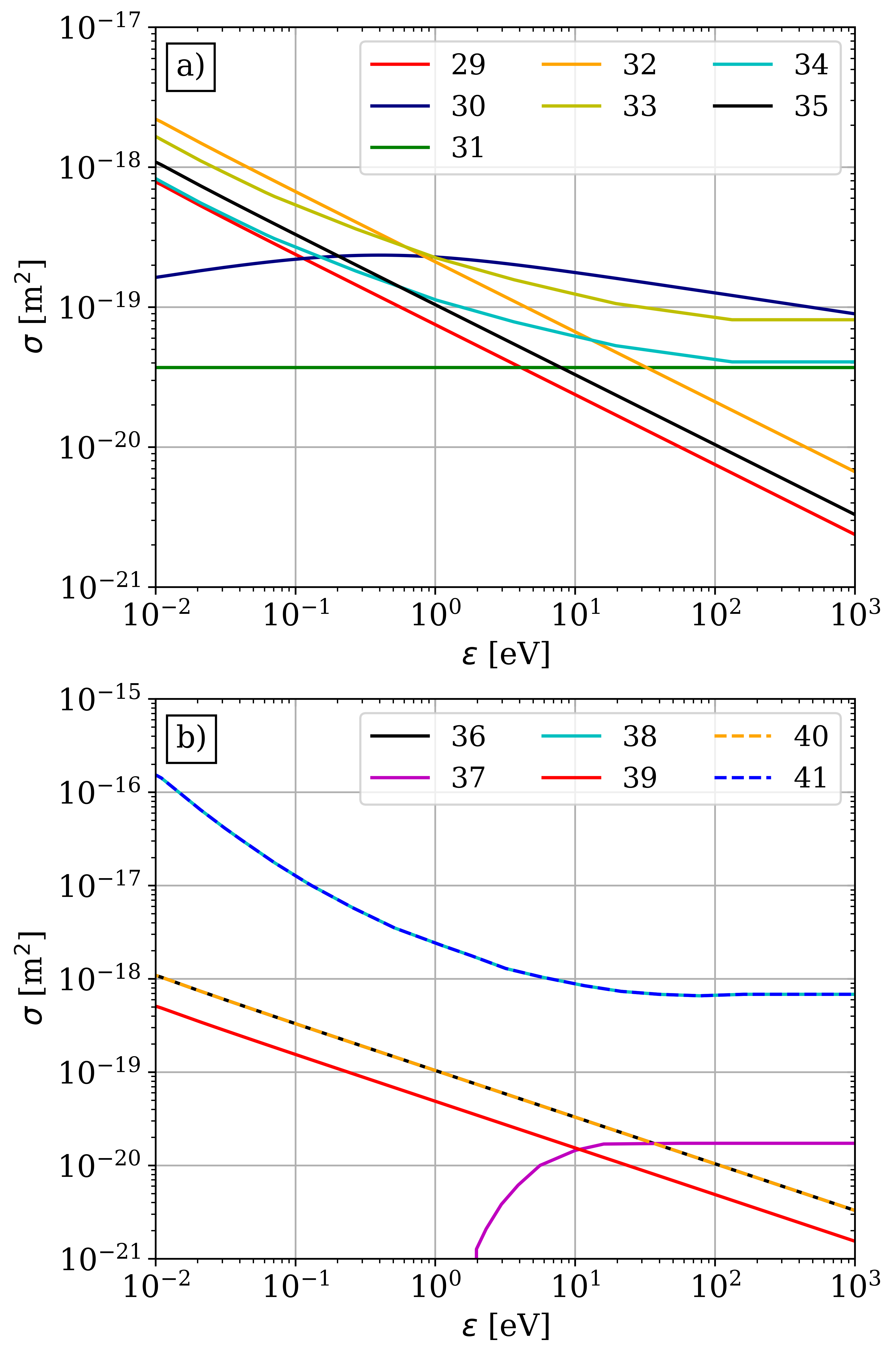}
	\caption{Cross sections of the collision processes for Ne$^+$, $\rm{O_2^+}$ and $\rm{O^-}$ ions, listed in table~\ref{table:Neon-Oxygen-mixture} (processes 29--41), as a function of the kinetic energy (considered in the center-of-mass frame) of the projectile: 
	(a) Isotropic and backward elastic scattering between Ne$^+$ ions and Ne atoms (processes 29 and 30, respectively), Ne$^+$ ions and O$_2$ molecules (processes 32 and 31, respectively), $\rm{O_2^+}$ ions and O$_2$ molecules (processes 33 and 34, respectively), and isotropic elastic scattering between $\rm{O_2^+}$ ions and Ne atoms (process 35).
	(b) Elastic scattering (process 36) and detachement (process 37) between $\rm{O^-}$ ions and O$_2$ molecules, mutual neutralization of $\rm{O^-}$ with $\rm{O_2^+}$ (process 38) and Ne$^+$ (process 41), associative detachment of $\rm{O^-}$ with $\rm{O_2 (a^1 \Delta_g)}$ (process 39) and isotropic elastic scattering with Ne atoms (process 40). Note that the cross sections overlap in case of processes 36 and 40, as well as in case of processes 38 and 41.}
	\label{fig:ion_collisions}
\end{figure}

\subsubsection*{Electron collisions:}
The collision processes between electrons and Ne atoms are elastic scattering (process 1), excitation (processes 2--10) and ionization (process 11). Nine electron impact Ne excitation processes are considered, including the Ne~$\rm{3p_0}$ excitation from the ground state (process 10), whose population dynamics is captured experimentally by the PROES measurements. The cross sections for the e$^-$ + Ne collisions are adopted from the Biagi-v7.1 dataset \cite{Biagi7.1} and are plotted in figure \ref{fig:electron_collisions}(a).

Collisions between electrons and $\rm{O_2}$ molecules include elastic scattering (process 12), rotational and vibrational excitation (processes 13--17), metastable excitation (processes 18 and 19), dissociative attachment (process 20), electronic excitation (process 21), dissociation (processes 22-24), ionization (process 25), and dissociative excitation (process 26). Electron impact detachment of $\rm{O^-}$ ions (process 27) and dissociative recombination of $\rm{O_2^+}$ ions (process 28) are also considered. The e$^-$ + O$_2$ process list and cross sections adopted are largely based on the ``xpdp1'' set \cite{vahedi1995monte}. However, a few changes have been introduced \cite{Derzsi_2015}: we (i) replace the elastic collision (process 12) cross section with the elastic momentum transfer cross section of Biagi \cite{Biagi8.9} and use, accordingly, isotropic electron scattering, (ii) replace the original ``xpdp1'' cross sections for the ionization (process 25), electron impact detachment (process 27) and dissociative recombination (process 28) with those recommended by Gudmundsson \cite{gudmundsson2013benchmark}. The cross sections for the electron impact collisions with oxygen species are plotted in figure~\ref{fig:electron_collisions}(b) (processes 12--17) and figure~\ref{fig:electron_collisions}(c) (processes 18--28).

\subsubsection*{Ion collisions:}
For Ne$^+$ ions, elastic collisions with Ne atoms and O$_2$ molecules are taken into account. In case of Ne$^+$ + Ne collisions, an isotropic channel (process 29) and a backward scattering channel (process 30) is considered and the cross sections are taken from \cite{PhepsNeonJILA} in the functional forms $\sigma_{29}(\varepsilon_{\rm com})= 1.06\cdot 10^{-19} \varepsilon_{\rm com}^{-0.5}$~m$^2$ and $\sigma_{30}(\varepsilon_{\rm com})= 2.8\cdot 10^{-19} \varepsilon_{\rm com}^{-0.15} (1+0.8/\varepsilon_{\rm com})^{-0.3}$~m$^2$, where $\varepsilon_{\rm com}$ is the value of the kinetic energy in the ion--atom center-of-mass reference frame in units of eV.

As Ne$^+$ + O$_2$ collisions, charge transfer (process 31) and isotropic elastic scattering (process 32) are included in the model. Process 31 is a non-resonant charge transfer reaction that is possible as the ionization potential of Ne is higher than that of O$_2$. The rate coefficient of this reaction at thermal energies was reported in \cite{adams1972thermal}. Schlumbohm \cite{schlumbohm1969dissoziativer} reported a measurement of this cross section in the energy range of 3-200 eV, and found it to be between 3.5 $\times 10^{-20}$~m$^{2}$ and 4 $\times 10^{-20}$~m$^{2}$. We use a cross section of constant 3.7$\times 10^{-20}$~m$^{2}$ for this process. 
Process 32 is treated with its Langevin cross section:
\begin{equation}
    \sigma_{\rm L} = \sqrt{\frac{\alpha \pi e^2}{\epsilon_0 \mu}}\frac{1}{g} = \sqrt{\frac{\alpha \pi e}{2 \epsilon_0}}\varepsilon_{\rm com}^{-1/2},
\end{equation}
where $\mu$ is the reduced mass, $g$ is the relative velocity and $\varepsilon_{\rm com}$ is kinetic energy in the centre-of-mass frame in units of eV. The polarizibility values are:
$\alpha$(O$_2$)~=~1.562~$\times~10^{-30}$~m$^{3}$ and 
$\alpha$(Ne)~=~0.381~$\times~10^{-30}$~m$^{3}$~\cite{Olney1997}. The cross sections of the collision processes used for Ne$^+$ ions are shown in figure~\ref{fig:ion_collisions}(a).

The $\rm{O_2^+}$ + $\rm{O_2}$ collisions can result in charge exchange or isotropic elastic scattering (processes 33 and 34, respectively). The cross sections for these processes are taken from Gudmundsson’s work \cite{gudmundsson2013benchmark}. Between O$_2^+$ ions and Ne atoms isotropic elastic scattering (process 35) is considered. Similarly to process 32, this process is treated with its Langevin cross section. The cross sections of the collision processes used for $\rm{O_2^+}$ ions are included in figure~\ref{fig:ion_collisions}(a).

For $\rm{O^-}$ + $\rm{O_2}$ collisions elastic scattering (process 36) and detachment (process 37) are considered. Mutual neutralization of $\rm{O^-}$ and $\rm{O_2^+}$ ions (process 38) and associative detachment between $\rm{O^-}$ ions and $\rm{O_2 (a^1 \Delta_g)}$ singlet delta oxygen molecules (process 39) are also included in the model. The metastable $\rm{O_2 (a^1 \Delta_g)}$ singlet delta oxygen molecules are not traced in the simulation, however, their density is calculated self-consistently based on the balance of their creation, de-excitation, and diffusive transport as discussed in section~\ref{sec:Metastable} below. The cross sections of processes 36--38 are from \cite{gudmundsson2013benchmark}, for process 39 the cross section is taken from \cite{bronold2007radio}. 
As ``cross-process''  between O$^-$ ions and Ne species, isotropic elastic scattering with Ne atoms (process 40) and mutual neutralisation with Ne$^+$ ions (process 41) is considered. Process 40 is treated with its Langevin cross section. Process 41 is treated in a similar way as the mutual neutralisation O$^-$~+~O$_2^+$~$\longrightarrow$ O~+~O$_2$ (process 38). Due to the lack of cross section data for O$^-$~+~Ne$^+$~$\longrightarrow$ O~+~Ne neutralization, we use the same cross section for process 41 as that used for process 38. The cross sections of the collision processes used for $\rm{O^-}$ ions are plotted in figure~\ref{fig:ion_collisions}(b).

\subsubsection*{Fast neutral collisions:}

\begin{figure}[ht]
	\centering
	\includegraphics[width=0.5\linewidth]{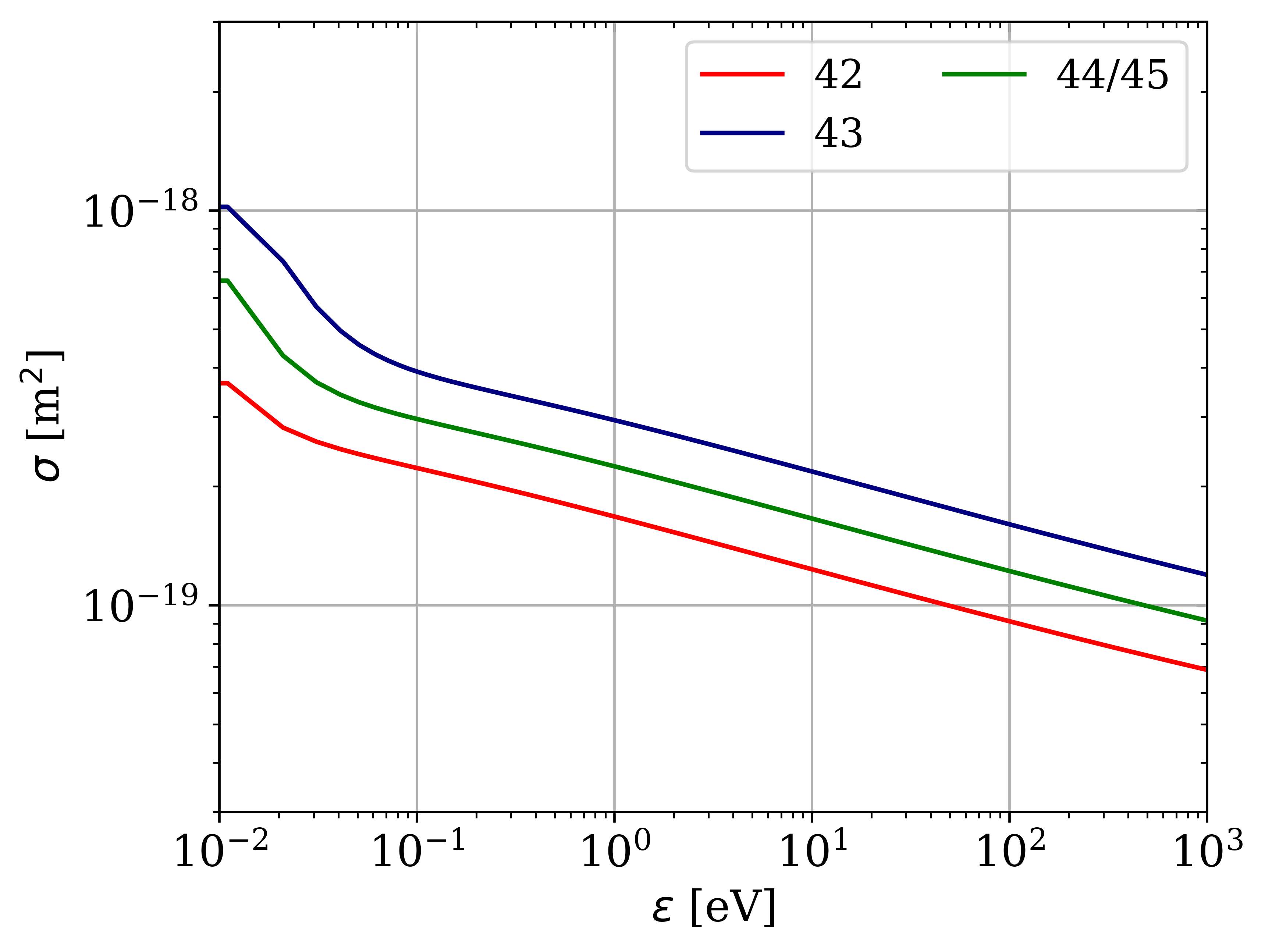}
	\caption{Cross sections of the collision processes for fast Ne atoms (Ne$^{\rm f}$) and fast O$_2$ molecules (O$_2^{\rm f}$), listed in table~\ref{table:Neon-Oxygen-mixture} (processes 42--45), as a function of the kinetic energy (considered in the center-of-mass frame) of the projectile: elastic scattering between Ne$^{\rm f}$ and Ne atoms/O$_2$ molecules (process 42/44) and elastic scattering between O$_2^{\rm f}$ and O$_2$ molecules/Ne atoms (process 43/45). Note, that the cross section of processes 44 and 45 are the same (see text).}
	\label{fig:fastneutral_collision}
\end{figure}

A background gas atom or molecule acquiring a kinetic energy greater than a threshold energy of $E_{\rm thr} = 9 \cdot \tfrac{3}{2} k_{\rm B} T$ after a collision is considered to be a fast neutral, and is traced in the simulation until its energy drops below $E_{\rm thr}$. Thermalization of fast neutrals due to collisions with atoms/molecules of the background gas contributes to the heating of the gas (see section~\ref{sec:Heating}). Elastic collisions of fast Ne atoms (Ne$^{\rm f}$) and fast O$_2$ molecules (O$_2^{\rm f}$) with background O$_2$ molecules and Ne atoms (processes 42–45) are considered in the model in the following way: (i) the scattering cross sections are derived from $\sigma_{\rm V}$, the viscosity cross section and (ii) isotropic scattering in the COM frame is assumed. 
As reliable cross section data for Ne$^{\rm f}$ and O$_2^{\rm f}$ for neon-oxygen mixture are not available in the literature, the cross sections are calculated for processes 42--45 based on the pair-potential between the particles, for which the Lennard-Jones (LJ) type is assumed:
\begin{equation}
    \phi(r) = 4 \epsilon_{\rm LJ} \biggl[ \biggl( \frac{r}{\sigma_{\rm LJ}} \biggr)^{-12} - \biggl( \frac{r}{\sigma_{\rm LJ}} \biggr)^{-6} \biggr],
\end{equation}
where the energy $\epsilon_{\rm LJ}$ and the characteristic length $\sigma_{\rm LJ}$ can be found in tables, e.g. in \cite{pratt1979pulverized}. The calculation of the cross section starts with computing the scattering angle $\chi$ as a function of the impact parameter $b$. This is done as described in \cite{matsumoto1991comparison}. Having obtained the $\chi(b)$ function for a given LJ potential, the viscosity cross section is computed as \cite{khrapak2014accurate}:
\begin{equation}
    \sigma_{\rm V} = \int_0^\infty \sin^2(\chi) b~{\rm d}b.
\end{equation}
The $\chi(b)$ function is computed for a set of energies (in the COM frame) and the above integral is evaluated for all these energy values, resulting in the energy-dependent elastic scattering cross section $\sigma_{\rm fast}(\varepsilon) = \frac{3}{2} \sigma_{\rm V}(\varepsilon)$. 
To obtain the LJ parameters for the Ne--O$_2$ "cross collisions", the Lorentz-Berthelot combining rule is used~\cite{schnabel2007unlike}:
\begin{eqnarray}
\sigma_{\rm LJ}^{ab} = \frac{\sigma_{\rm LJ}^{aa}+\sigma_{\rm LJ}^{bb}}{2},\\
\epsilon_{\rm LJ}^{ab} = \sqrt{\epsilon_{\rm LJ}^{aa}\, \epsilon_{\rm LJ}^{bb}},
\end{eqnarray}
where $a$ and $b$ denote the interacting species.
The $a \leftrightarrow b$ symmetry of these equations results in the same cross sections for 
Ne$^{\rm f}$+O$_{2}$ and O$_2^{\rm f}$+Ne collisions.
These computed cross sections (for processes 42–45) are plotted in figure~\ref{fig:fastneutral_collision}. The calculation of the cross sections of fast neutrals was validated by comparing our results for argon gas with those given in \cite{khrapak2014accurate}.


\subsection{Surface processes}\label{sec:SurfaceProcesses}

The surface processes taken into account in the discharge model are electron reflection, secondary electron emission (SEE) induced by O$_2^+$ ions and Ne$^+$ ions, and surface quenching of $\rm{O_2 (a^1 \Delta_g)}$ metastable molecules. Constant surface coefficients are specified for these processes. At the electrode surfaces, electrons are elastically reflected with a probability of $\eta_{\rm{e}}=0.7$. This value is adopted based on the findings of Schulenberg \emph{et al.} \cite{Schulenberg21}. For O$_2^+$ ions, a SEE coefficient of 0.015 is considered, while for Ne$^+$ ions the SEE coefficient is set to 0.1~\cite{Magnusson2020}. 
The value of the surface quenching probability (surface destruction probability), which controls the loss rate of $\rm{O_2 (a^1 \Delta_g)}$ metastable molecules, is set to $\alpha = 8 \cdot 10^{-4}$ in this study (see discussion in section~\ref{sec:Metastable}). 

\subsection{Gas heating}\label{sec:Heating}

The collision frequency of elementary processes is strongly dependent on the local gas temperature (through the dependence on the background gas density). In order to cover the wide range of discharge parameters properly, gas heating needs to be included in the model. The coupling of the input RF electric power to the gas heating is realised by two processes: (i) in the gas phase, ions are accelerated by the electric field and energy is transferred to the background gas atoms/molecules through collisions, mediated by fast neutrals, and (ii) at the surfaces, the flux of particles interacting with the electrodes causes them to heat up. 
\subsubsection*{Heat transfer:}
The thermal power density, $P(x)$, accumulates the excess energy transferred to the background gas, originating from the particles accelerated by the electric field, per unit time and volume~\cite{Serikov97}. It acts as the source term in the steady-state heat equation, that is included in the model to calculate the equilibrium temperature distribution in the discharge gap:
\begin{equation}
    \frac{\partial^2 T_\text{gas}(x)}{\partial x^2}=-\frac{1}{\kappa} P(x),
    \label{eqn:heat}
\end{equation}
were $\kappa$ is the thermal conductivity specific to the gas mixture.

Analogously to the linear particle shape function in the PIC method, if a collision with the background gas happens within the $i$-th grid cell, it contributes to $P(x)$ at the neighbouring grid points in the following way:
\begin{eqnarray}\label{eqn:dPi}
    {\rm d} P_{i} &=& \frac{\Delta E}{A \Delta x \Delta t_{\rm{H}}} \left( (i + 1) - \frac{x}{\Delta x} \right), \\
    {\rm d} P_{i+1} &=& \frac{\Delta E}{A \Delta x \Delta t_{\rm{H}}} \left(\frac{x}{\Delta x}-i\right), \nonumber
\end{eqnarray}
where $A$ is the surface of the electrodes, $\Delta x$ is the length of the grid cell normal to the electrodes, and $\Delta t_{\rm{H}}$ is the time of data collection (several RF cycles). $\Delta E$ is the energy gain of a background gas atom/molecule due to the collision. Collisions of ions, as well as fast neutrals with the background gas, contribute to $P(x)$. When thermalized, the neutrals deposit all their energy to the $P(x)$ thermal power.

\subsubsection*{Boundary condition:} 
In order to solve eq.~(\ref{eqn:heat}), the boundary values of the temperature distribution have to be defined. In regimes with Knudsen number (Kn) close but below unity the ``slip flow'' approximation can be applied~\cite{Gombosi}. The Knudsen number is defined as $\text{Kn} = l/\Lambda$, where $l$ is the molecular mean free path and $\Lambda$ is the characteristic size of the system, i.e. the length of the discharge gap in our case. The resulting boundary condition is of the third kind, introducing a temperature jump at the electrodes as described in~\cite{Sazhin97,Kennard}:
\begin{equation}
    T_\text{gas}(x=0/L) = T_{\rm wall} + \lambda \left|\frac{\partial T_\text{gas}}{\partial x} \right|_{x=0/L},
    \label{eqn:boundcond}
\end{equation}
where $T_{\rm wall}$ is the temperature of the walls (i.e. the electrodes). $\lambda$ is called the temperature jump distance, and is proportional to the mean free path perpendicular to the electrodes, and it depends on the gas mixture. For further details see~\cite{Sazhin97}. Note that eq.~(\ref{eqn:boundcond}) imposes conditions on the slope of the temperature distribution at the boundaries. To satisfy these, the following iterative method is used:
(i) Initially the temperature jump is neglected and wall temperatures ($T_{\rm wall}$) are used as boundary values for the Thomas algorithm which solves eq.~(\ref{eqn:heat}).
(ii) Since the resulting temperature distribution does not satisfy eq.~(\ref{eqn:boundcond}), the right hand side of eq.~(\ref{eqn:boundcond}) is set as new boundary values to solve eq.~(\ref{eqn:heat}) again.
(iii) Step (ii) is repeated until condition (\ref{eqn:boundcond}) is fulfilled within pre-defined  tolerance limits. The convergence of this method was demonstrated in~\cite{Sazhin97}.

\begin{figure}[ht]
	\centering
	\includegraphics[width=0.6\linewidth]{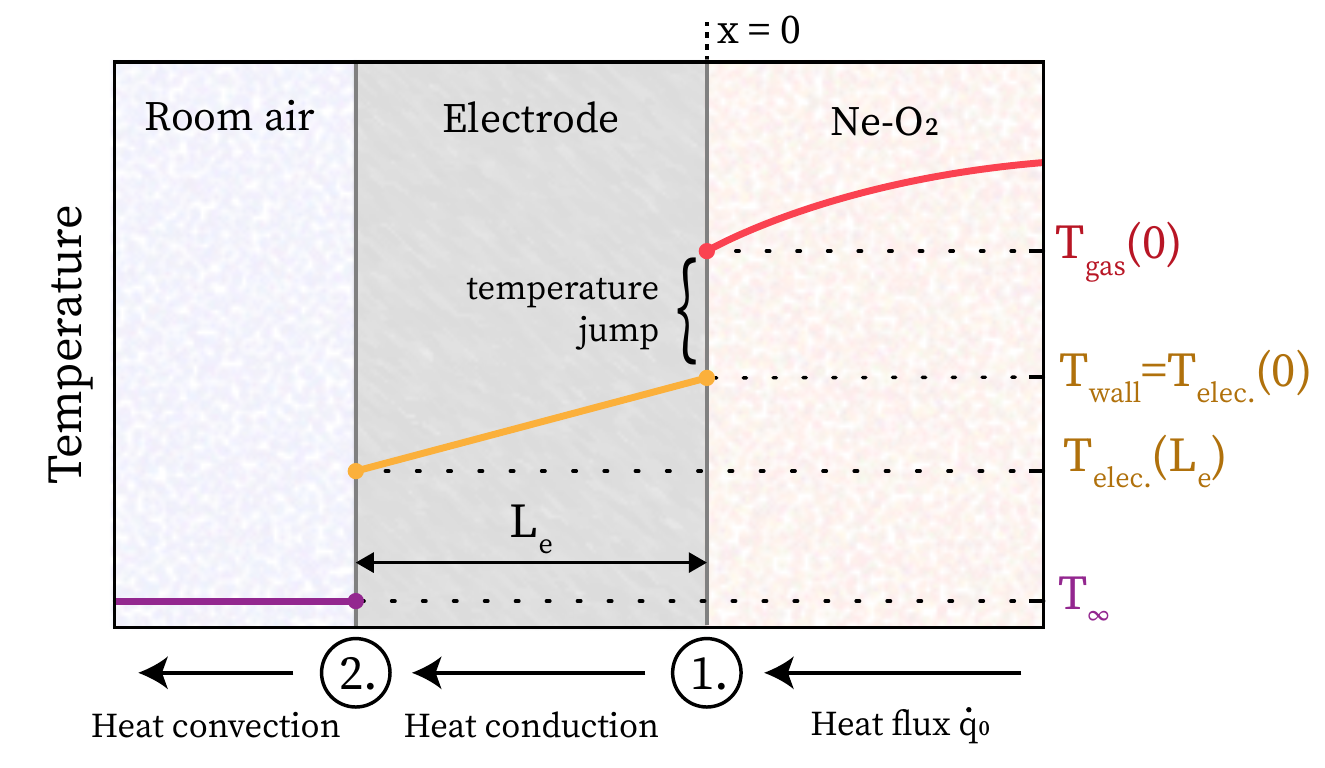}
	\caption{Simplified geometry assumed in the model for heat convection at the electrodes, showing the relevant quantities and a sketch of the temperature profile.}
	\label{fig:heat_convection}
\end{figure}

\subsubsection*{Heat convection at the electrodes:}
The heating of the electrode surfaces is caused by inelastic collisions of plasma particles with the electrodes. In order to accurately describe the experimental system, this change of wall (electrode) temperature must be accounted for in the gas heating module, as it enters directly in eq.~(\ref{eqn:boundcond}). The equilibrium wall temperature is determined by the balance of heat influx at the plasma side, thermal conduction of the electrode assembly, and heat dissipation to the environment at the outer surface of the system. In our calculations, we consider a simplified geometry of the electrode arrangement, as compared to the experimental realisation (consisting of the electrodes, spacers, and flanges), which is shown in figure~\ref{fig:heat_convection}. This simplification is permissible because of the use of a calibration procedure for the effective heat conductivity of the setting, to be revealed below. Let $\dot{q}_0$ denote the heat flux originating from the discharge reaching the electrode and $T_\infty$ the room temperature. $k$ and $L_{\rm e}$ are the thermal conductivity and the width of the electrode, respectively. $h$ is the convection heat transfer coefficient of air. For the model system of figure~\ref{fig:heat_convection}, the steady-state homogeneous heat equation inside the electrode and its boundary conditions read:
\begin{eqnarray}\label{eqn:heatwall}
    \frac{\text{d}^2T_\text{elec.}(x)}{\text{d}x^2}&=&0 {\rm~~~~for~}0<x<L_{\rm e}, \\
    -k\frac{\text{d}T_\text{elec.}(0)}{\text{d}x} &=&\dot{q_0}, 
    \label{eqn:boundcond_flux} \\
    -k\frac{\text{d}T_\text{elec.}(L_{\rm e})}{\text{d}x} &=& h\left[ T_\text{elec.}(L_{\rm e})-T_\infty \right],
    \label{eqn:boundcond_conv}
\end{eqnarray}
where $x=0$ represents the plasma side and $x=L_{\rm e}$ the external surface of the electrode assembly. Eqs.~(\ref{eqn:boundcond_flux}) and (\ref{eqn:boundcond_conv}) are boundary conditions for boundaries at point 1. and 2. in figure~\ref{fig:heat_convection}. The solution of eq.~(\ref{eqn:heatwall}) yields the wall temperature in terms of the heat flux from the plasma, assuming equal $\dot{q}_0$, $k$, and $L_{\rm e}$ values for both electrodes:
\begin{equation}\label{eqn:Twall}
    T_\text{wall} = T_\text{elec.}(0) = T_\infty + \dot{q}_0\left( \frac{L_{\rm e}}{k} + \frac{1}{h}\right).
\end{equation}

The value of $\dot{q}_0$ could, in principle, be calculated tracing individual particle--surface interactions, however, an even simpler and more precise estimation can be achieved by assuming that, in equilibrium, all electric power absorbed by the plasma (charged plasma particles only) from the RF field becomes finally absorbed by the electrodes and is dissipated in form of heat. This means that the heat flux reaching one electrode is the time average ($\langle...\rangle$) of the half of the total electrical input power per unit area:
\begin{equation}
    \dot{q}_0=\frac{1}{2A}\left<\sum_i q_i W_i ({\bf E}_i \cdot {\bf v}_i)\right>,
\end{equation}
where the sum runs over all charged particles $i$ with charge $q_i$, weight factor $W_i$, velocity ${\bf v}_i$ and electric field at the particles' position ${\bf E}_i$. Having calculated this quantity, the wall temperature can readily be obtained from eq.~(\ref{eqn:Twall}), if $(L_{\rm e}/k+1/h)$ is known. While this factor (that consists of three components, each characterizing different aspects of the overall heat dissipation) could be approximated based on the exact dimensions and material properties of the electrode constructions, we have chosen an alternative way to determine it, using a calibration technique. Tunable diode laser absorption spectroscopy (TDLAS) measurements were performed in pure Ar discharges in the same experimental system. The details of this technique are described in~\cite{Schulenberg21}. The measured gas temperature was then compared with PIC/MCC results obtained for different values of the $(L_{\rm e}/k+1/h)$ coefficient, and its value giving the best agreement was determined and used throughout this study.

\subsection{Oxygen metastable balance}\label{sec:Metastable}

Previous studies on oxygen CCPs have shown that at low pressures the discharge contains more O$^-$ ions than electrons, thus it is electronegative~\cite{Derzsi_2016,Gudmundsson_2017,Vass_2020}, which is quantified by the global parameter $\beta = \langle n_{\rm O^-}\rangle / \langle n_{\rm e}\rangle$, i.e., the global electronegativity of the discharge. The O$^-$ ion density is determined by the balance of its creation (by means of electron impact dissociative attachment of ${\rm O_2}$ molecules, process 20 in table~\ref{table:Neon-Oxygen-mixture}) and the primary loss in the gas phase due to associative detachment in collisions with ${\rm O_2(a^1\Delta_g)}$ metastable molecules (process 39 in table~\ref{table:Neon-Oxygen-mixture}). In order to accurately compute the O$^-$ ion density distribution, it is important to determine the density of the ${\rm O_2(a^1\Delta_g)}$ metastable species self-consistently. This is done by modeling the balance of creation, transport, and surface de-excitation of the ${\rm O_2(a^1\Delta_g)}$ molecules, which are incorporated as continuum species, described by the stationary diffusion equation
\begin{equation}\label{eqn:diff}
    D\frac{\partial^2 n_{\rm m}(x)}{\partial x^2}=S_{\rm m}(x),
\end{equation}
where $D$ is the diffusion coefficient, $n_{\rm m}(x)$ and $S_{\rm m}(x)$ are the density and time averaged (over several RF cycles) source rate distributions of the ${\rm O_2(a^1\Delta_g)}$ metastables. Like in the gas temperature calculation module, here we aim only for the stationary solution of the transport equation and assume that it does not depend on the actual time-evolution, which, on the other hand, can be significantly slower than the relaxation of other plasma parameters. The source term in eq.~(\ref{eqn:diff}) is calculated analogously to the thermal power input in eq.~(\ref{eqn:dPi}) but with the microscopic contribution of process 18 in table~\ref{table:Neon-Oxygen-mixture} for every electron in each time step. The contribution of gas phase losses of metastables (through processes 39 in table~\ref{table:Neon-Oxygen-mixture}) is not considered in the metastable density balance calculation. The validity of this simplification is supported by the collision statistical analysis, showing that the average probability of associative detachment is $<1$\% of the probability of metastable excitation for all conditions.

To obtain proper pressure, temperature, and gas composition dependent diffusion coefficients we utilize the predictions of the Chapman-Enskog theory~\cite{Chapman1939,Winn1950}, resulting in the general form
\begin{equation}\label{eqn:DpT}
    D\approx D_0 \frac{p_0}{p}\left(\frac{T}{T_0}\right)^{3/2},
\end{equation}
where $D_0$ is the gas composition dependent diffusion coefficient at a reference pressure $p_0=1$~Pa and reference temperature $T_0=300$~K. The values of $D_0$ for O$_2$ molecules in pure O$_2$ and Ne gases are estimated as $D_0^{\rm O_2} = 2.07~{\rm m}^2/{\rm s}$ and $D_0^{\rm Ne} = 3.29~{\rm m}^2/{\rm s}$. Values for arbitrary mixing ratios are calculated by linear interpolation between these two extremes.

The dominant loss process of ${\rm O_2(a^1\Delta_g)}$ metastables is the de-excitation at the electrodes, which is incorporated in the boundary conditions. Similarly to the gas heating calculation, due to the low pressure condition, where the collision mean free path is comparable to the characteristic size of the system, the boundary condition is of the third kind, relating the boundary value with its gradient as
\begin{equation}\label{eqn:BCdiff} 
    \frac{n_{\rm m}(x=0/L)}{\kappa} = \left|\frac{\partial n_{\rm m}}{\partial x} \right|_{x=0/L},
\end{equation}
with $\kappa = l(2-\alpha)/(\sqrt{3}\alpha)$, where $l$ is the collision mean free path of O$_2$ molecules, and $\alpha = 8 \cdot 10^{-4}$ is the surface destruction (de-excitation or quenching) probability. The mean free path is calculated from the diffusion coefficient using the formula derived from the kinetic theory of gases~\cite{StatPhys}
\begin{equation}\label{eqn:MFP}
    l = \frac{4}{\sqrt{3\pi}}\left(\frac{m}{3k_{\rm B}T}\right)^{1/2}D.
\end{equation}

With all terms evaluated eq.~(\ref{eqn:diff}) is solved using the Thomas tridiagonal algorithm, the metastable density values at the boundaries $n_{\rm m}(x=0/L)$ are adjusted, and the solution is iterated until the boundary conditions (\ref{eqn:BCdiff}) are met. 

\subsection{Electron power absorption}
\label{sec:EPowerAbsorption}

The characteristics of the electron power absorption have been investigated using the Boltzmann term method. This method, which provides spatio-temporally resolved information about the electric field and the power absorbed by the electrons, has been thoroughly described in earlier works \cite{Schulze2018_Boltzmann,Vass_2020,Vass_2021}. Therefore, here only a short overview is given. The Boltzmann term analysis is based on the electron momentum balance equation, which can be rearranged for the electric field, which then can be divided into various terms, given by $E_{\rm tot}=E_{\rm in}+E_{\nabla p}+E_{\rm Ohm}$ \cite{Vass_2021}, where
\begin{align}\label{Eterm}
    E_{\rm in}&=-\frac{m_{\rm e}}{n_{\rm e}e}\left[\frac{\partial}{\partial t}(n_{\rm e}u_{\rm e})+\frac{\partial}{\partial x}(n_{\rm e}u_{\rm e}^2)\right],\nonumber \\
    E_{ \nabla p}&= - \frac{1}{n_{\rm e}e} \frac{\partial}{\partial x} p_{\parallel}, \nonumber \\	
    E_{\rm Ohm}&=-\frac{\Pi_{\rm c}}{n_{\rm e}e}. 
\end{align} 
Here $m_{\rm e}$ and $e$ are the electron mass and charge, respectively, $n_{\rm e}$ is the electron density, $u_{\rm e}$ is the mean velocity, $\Pi_{\rm c}$ is the electron momentum loss and $p_{\parallel}$ denotes the diagonal element of the pressure tensor. Each of these electric field terms corresponds to a distinct physical mechanism: $E_{\rm in}$ is the electric field term originating from inertial effects; $E_{\rm Ohm}$, the Ohmic electric field, is a consequence of electrons colliding with the particles of the background gas, and $E_{\nabla p}$ is due to pressure effects. This electric field term can be split into two additional terms, according to $E_{\nabla p}=E_{\nabla n}+E_{\nabla T}$, where
\begin{align}\label{Egradp}
E_{\nabla n}&=-\frac{T_{\parallel}}{n_{\rm e}e}\frac{\partial n_{\rm e}}{\partial x}, \nonumber \\
E_{\nabla T}&=-\frac{1}{e}\frac{\partial T_{\parallel}}{\partial x},
\end{align} 
where $T_\parallel = p_\parallel/n_{\rm e}$ is the parallel electron temperature. $E_{\nabla n}$ is, in regions where quasineutrality holds, identical to the ``classical ambipolar'' electric field \cite{Schulze2014_ambipolar} and $E_{\nabla T}$ originates from the spatial gradient of the electron temperature. From the electric field terms the corresponding power absorption terms can be easily calculated by multiplying each of these terms with the spatio-temporally resolved electron conduction current density, $j_{\rm c}$.

\subsection{Simulation parameters} \label{sec:Sim_settings}


Table~\ref{tab:Sim_params} lists the most important physical and numerical parameters of the simulations grouped according to the gas pressure. The physical parameters were chosen to cover a low to intermediate pressure regime and a wide mixing range.
The numerical parameters were chosen to ensure the fulfillment of the usual stability and accuracy requirements imposed on the explicit electrostatic PIC/MCC scheme. For a recent detailed discussion on these criteria and the effect of different numerical parameters on the simulation results see e.g.~\cite{NumericPIC}. 

\begin{table*}[h]
\centering
\caption{Physical and numerical parameters of the discharges studied.}
\resizebox{\textwidth}{!}{\begin{tabular}{lc|cccccc}
    \hline
    \multicolumn{2}{l|}{\it \textbf{Physical parameters:}} & & & & & & \\
    gas pressure               & $p$          & 15 Pa    & 31 Pa    & 62 Pa   & 125 Pa  & 250 Pa  & 500 Pa  \\
    O$_2$ concentration range  & [\%]         & 50--90   & 10--90   & 10--90  & 10--90  & 10--90  & 10--70  \\ 
    electrode separation       & $L$          & \multicolumn{6}{c}{2.5 cm}  \\
    environment temperature    & $T_\infty$   & \multicolumn{6}{c}{295 K}   \\ 
    peak-to-peak voltage       & $V_{\rm pp}$ & \multicolumn{6}{c}{350 V}   \\ 
    frequency                  & $f$          & \multicolumn{6}{c}{10 MHz~~~($T_{\rm RF}=1/f=100$~ns)} \\ 
    \hline
    \multicolumn{2}{l|}{\it \textbf{Numerical parameters:}} & & & & & & \\
    cell size            & $\Delta x$   & \multicolumn{6}{c}{$L/511$}   \\
    nominal electrode area  & $A$       & \multicolumn{6}{c}{1 cm$^2$}  \\
    time step size       & $\Delta t$   & $T_{\rm RF}/6000$ & $T_{\rm RF}/6000$ & $T_{\rm RF}/8000$ & $T_{\rm RF}/12000$ & $T_{\rm RF}/20000$ & $T_{\rm RF}/30000$ \\
    electron \& Ne$^+$ ion weight factor  & $W_{\rm e}$     & 1\,500   & 2\,500   & 4\,000   & 4\,000   & 8\,000   & 10\,000  \\
    O$_2^+$ \& O$^-$ ion weight factor    & $W_{\rm O}$     & 60\,000  & 50\,000  & 40\,000  & 40\,000  & 64\,000  & 50\,000  \\
    O$_2^{\rm f}$ molecule weight factor  & $W_{\rm O_2^f}$ & 120\,000 & 75\,000  & 40\,000  & 12\,000  & 8\,000   & 50\,000  \\
    Ne$^{\rm f}$ atom weight factor       & $W_{\rm Ne^f}$  & 60\,000  & 20\,000  & 12\,000  & 4\,000   & 8\,000   & 10\,000  \\
    \hline
\end{tabular}}
\label{tab:Sim_params}
\end{table*}

Further parameters specific to surface processes, heat conductivity, and metastable molecule diffusion are introduced in the previous sections.

\section{Results and discussion}\label{sec:Results}

\subsection{Experimental results}

\begin{figure}[ht]
	\centering
	\includegraphics[width=0.95\linewidth]{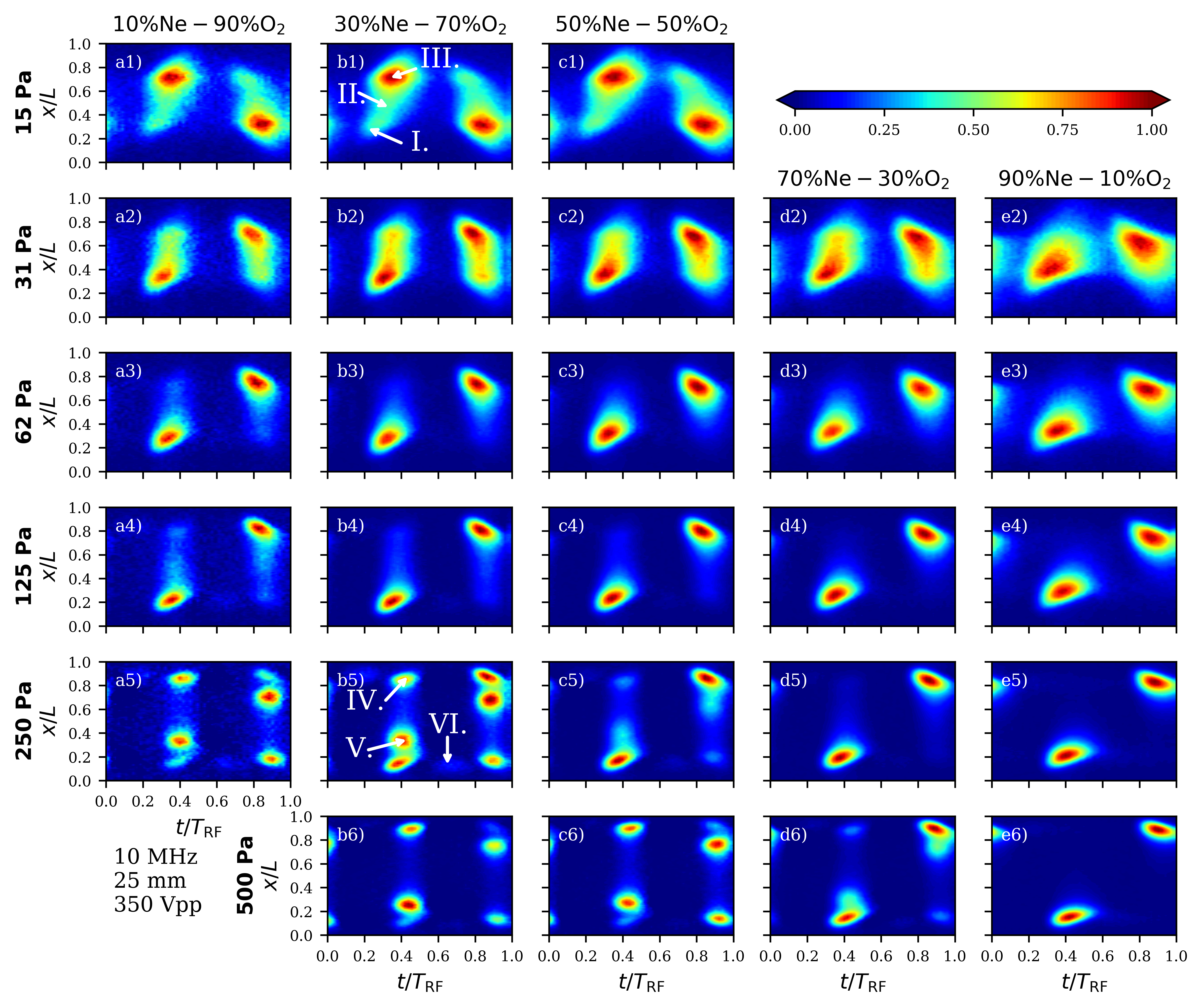}
	\caption{Spatio-temporal plots of the electron impact excitation rate from the ground state into the Ne~$\rm{3p_0}$ state measured by PROES [a.u.] in Ne-O$_2$ CCPs at different neutral gas pressures: 15~Pa (first row), 31 Pa (second row), 62~Pa (third row), 125~Pa (fourth row), 250~Pa (fifth row) and 500~Pa (sixth row), for different neon/oxygen concentration of the background gas mixture: 10\%~Ne--90\%~O$_2$ (first column), 30\%~Ne--70\%~O$_2$ (second column), 50\%~Ne--50\%~O$_2$ (third column), 70\%~Ne--30\%~O$_2$ (fourth column), 90\%~Ne--10\%~O$_2$ (fifth column). 
	The powered electrode is located at $x/L=0$, while the grounded electrode is at $x/L=1$. The labels I.-VI. in panels (b1) and (b5) indicate the different excitation features explained in the text.
	Discharge conditions: $L=2.5$~cm, $f=10$~MHz, $V_{\rm{pp}} = 350$~V.}
	\label{fig:exp}
\end{figure}

Figure~\ref{fig:exp} shows the spatio-temporal distribution of the electron impact excitation rate from the ground state into the Ne~$\rm{3p_0}$ state  measured by PROES at six different pressures: 15~Pa (first row), 31 Pa (second row), 62~Pa (third row), 125~Pa (fourth row), 250~Pa (fifth row) and 500~Pa (sixth row). At a given pressure (panels in a given row), results obtained for different neon/oxygen concentration of the background gas mixture (with decreasing O$_2$ concentration in panels from left to right) are shown: 10\%~Ne--90\%~O$_2$ (first column), 30\%~Ne--70\%~O$_2$ (second column), 50\%~Ne--50\%~O$_2$ (third column), 70\%~Ne--30\%~O$_2$ (fourth column), 90\%~Ne--10\%~O$_2$ (fifth column). All panels of figure~\ref{fig:exp} cover one RF period on the horizontal axes, and the vertical axes show the distance from the powered electrode.

At the lowest pressure of 15~Pa (first row), excitation at both the expanding and the collapsing sheath edges can be observed, as well as in the bulk region. The strongest excitation is found at the bulk side of the collapsing sheath edge at both electrodes. At this pressure, measurements have been performed for O$_2$ concentrations between 50\% and 90\% in the background gas mixture. For all cases, the spatio-temporal distribution of the excitation rate exhibits similar patterns (panels (a1)--(c1)), independently of the Ne/O$_2$ mixing ratio, suggesting hybrid $\alpha$-DA discharge operation mode (with dominant electron power absorption and excitation due to the DA-mode). The main excitation patterns observed at this pressure are labeled in panel (b1): I. indicates the excitation peak at the expanding sheath edge ($\alpha$-peak), II. denotes the excitation in the central bulk region (drift feature), and III. signals the excitation at the collapsing sheath edge (ambipolar peak).

At 31~Pa (second row), for
O$_2$ concentrations decreasing from 90\% to 30\% (panels (a2)-(d2)), strong excitation at the expanding sheath edge ($\alpha$-peak, I.) and significant excitation in the bulk region (drift feature, II.) with a weak excitation peak at the collapsing sheath edge (ambipolar peak, III.) are found, suggesting a hybrid $\alpha$-DA discharge operation mode (with dominant $\alpha$-mode). By decreasing the O$_2$ concentration at this pressure, the $\alpha$-peak (I.) at the expanding sheath edge is enhanced. At the lowest O$_2$ concentration of 10\%, the ambipolar peak (III.) at the collapsing sheath edge vanishes and the excitation plot suggests discharge operation in pure $\alpha$-mode (see panel (e2)).

At 62~Pa (third row), the excitation is concentrated at the expanding sheath edges, a strong $\alpha$-peak (I.) is exhibited, suggesting a dominant $\alpha$-mode for all Ne-O$_2$ mixtures. Weak excitation in the bulk region (drift feature, II.) is visible only in the case of the highest O$_2$ concentration (see panel (a3)).

At 125~Pa (fourth row), similarly to the results obtained at 62~Pa, strong $\alpha$ excitation peaks (I.) at the expanding sheath edges can be observed for all Ne-O$_2$ mixtures. At this pressure, excitation in the bulk region (drift feature, II.) is also present, for O$_2$ concentrations above 50\% (see panels (a4)-(c4)). By increasing the O$_2$ concentration in the mixture, the excitation in the bulk gets more and more pronounced.

At 250~Pa (fifth row), the excitation at the expanding sheath edges ($\alpha$-peak, I.) is present for all Ne-O$_2$ mixtures. Up to 50\% O$_2$ concentration this is the dominant excitation pattern (see panels (c5)-(e5)). Starting from 50\% O$_2$ concentration, the excitation in the bulk region is enhanced with the increasing O$_2$ content of the mixture (see panels (a5)-(c5)), and two distinct excitation peaks develop in the bulk. For 10\%~Ne--90\%~O$_2$ (panel (a5)) and 30\%~Ne--70\%~O$_2$ (panel (b5)), the strongest excitation can be observed in these patterns, labeled as IV. and V. in panel (b5), in the bulk region. At the highest O$_2$ concentration, the excitation is dominated by these new features (panel (a5)), which are different from the excitation patterns characteristic of the DA-mode. At 250~Pa pressure, at high O$_2$ concentrations, at both electrodes, there are two distinct excitation peaks (I. and V.) close to each other at the bulk side of the expanding sheath edge and one excitation peak (IV.) at the collapsing sheath edge. Under these conditions, the spatio-temporal distribution of the electron impact excitation rate to the Ne~$\rm{3p_0}$ state indicates the presence of the $\alpha$-mode in combination with some electron power absorption dynamics resulting in the formation of the two bulk excitation patterns (IV. and V.), the origin of which will be clarified later. 
For high O$_2$ concentrations, weak excitation at the phase of maximum sheath expansion at both electrodes can be also observed (see panels (a5) and (b5)), which suggests the presence of the $\gamma$-mode. The corresponding excitation pattern is labeled as VI. in panel (b5).

At the highest pressure of 500~Pa (sixth row), measurements have been carried out for up to 70\% O$_2$ concentration. At the lowest O$_2$ concentration of 10\%, a single excitation peak is seen at both electrodes at the expanding sheath edge ($\alpha$-peak, I.), indicating discharge operation in pure $\alpha$-mode (panel (e6)). Excitation in the bulk and development of two additional excitation peaks (IV. and V.) is visible starting from 30\% O$_2$ concentration (see panels (b6)-(d6)). Compared to the 250~Pa case, at this high pressure, the development of patterns IV. and V. starts at a lower O$_2$ concentration (30\% vs. 50\% at 250~Pa). These two excitation patterns are intensified as the O$_2$ concentration is increased, clearly dominating the excitation at 70\% and 50\% O$_2$ concentrations (panels (b6) and (d6)), while only weak excitation in the close vicinity of the expanding sheath edge ($\alpha$-peak, I.) can be observed in these cases.

The observed mode transitions as a function of the pressure and O$_2$ admixture are related to changes of the electronegativity of the discharge, which will be discussed in detail later based on the simulation results.

In summary, at the lowest pressure, weak $\alpha$-peak, strong ambipolar peak, and weak drift features are found in the excitation rate. With increasing pressure, the $\alpha$-peak and the drift feature are enhanced, while the ambipolar peak is reduced. At intermediate pressures, the $\alpha$-peak is the dominant excitation pattern in all mixtures. Further increase of the pressure leads to the formation of two distinct excitation peaks at the edges of the bulk region, which dominate the excitation at high O$_2$ concentrations. At the highest pressure, these excitation peaks are enhanced and the development of these features is found also in mixtures with lower O$_2$ concentration. 

\subsection{Simulation results}

In order to reveal the physics behind the formation of the different excitation patterns observed experimentally by PROES in the wide pressure range and gas mixing range presented above, especially those found at high pressures and high O$_2$ concentrations in the bulk region, PIC/MCC simulations have been performed. The PIC/MCC simulations covered the whole parameter regime studied by PROES, i.e. pressures between 15~Pa and 500~Pa, and  Ne/O$_2$ concentrations between 10\% and 90\%, at 10~MHz and 350~V peak-to-peak voltage.

\begin{figure}[ht]
	\centering
	\includegraphics[width=0.95\linewidth]{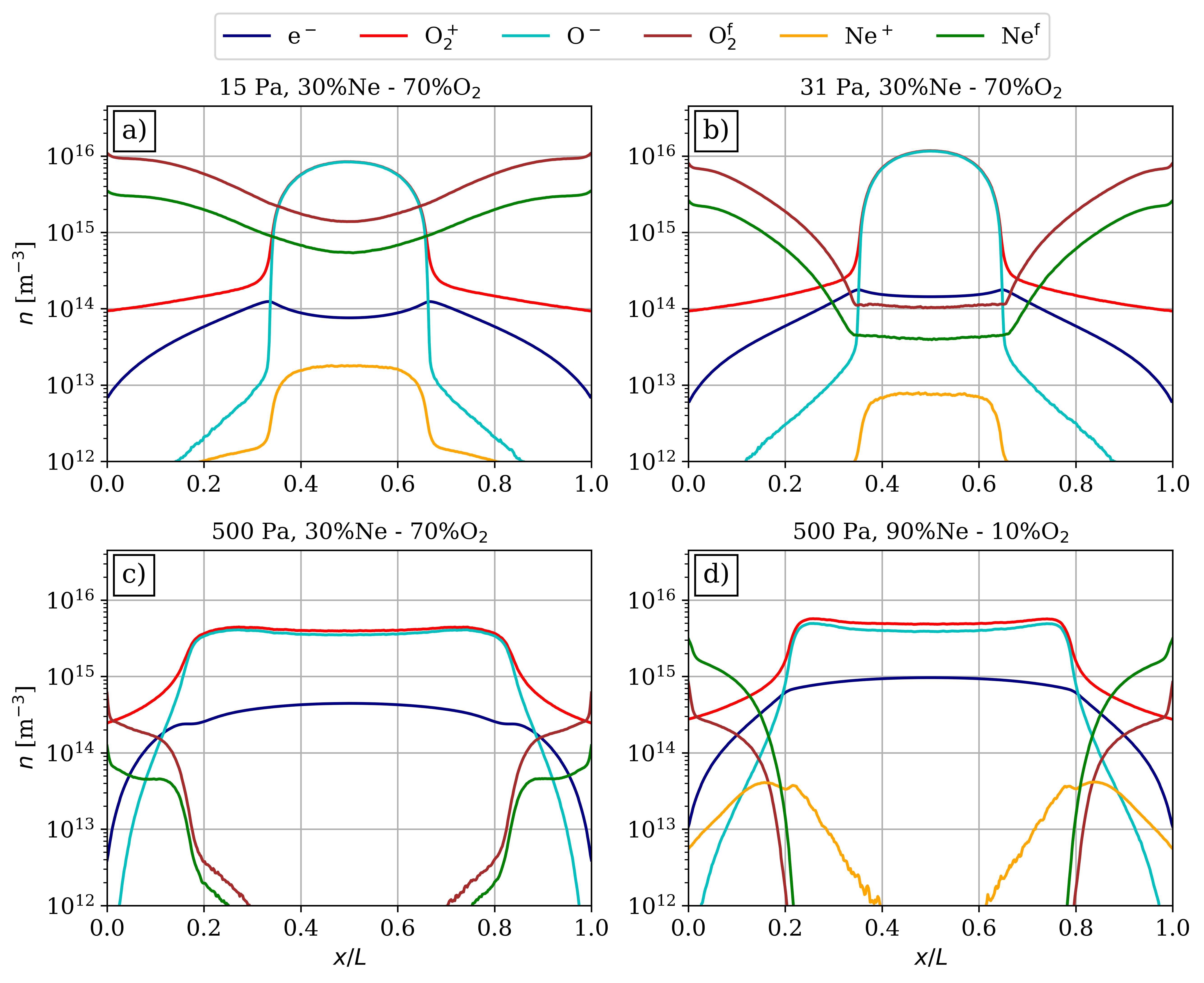}
	\caption{Time averaged particle density distributions obtained for different pressures and neon/oxygen concentrations of the background gas mixture: (a) 15~Pa, 30\%~Ne--70\%~O$_2$, (b) 31~Pa, 30\%~Ne--70\%~O$_2$, (c) 500~Pa, 30\%~Ne--70\%~O$_2$, and (d) 500~Pa, 90\%~Ne--10\%~O$_2$. Discharge conditions: $L=2.5$~cm, $f=10$~MHz, $V_{\rm{pp}} = 350$~V.}
	\label{fig:dens}
\end{figure}

\begin{figure}[ht]
	\centering
	\includegraphics[width=0.5\linewidth]{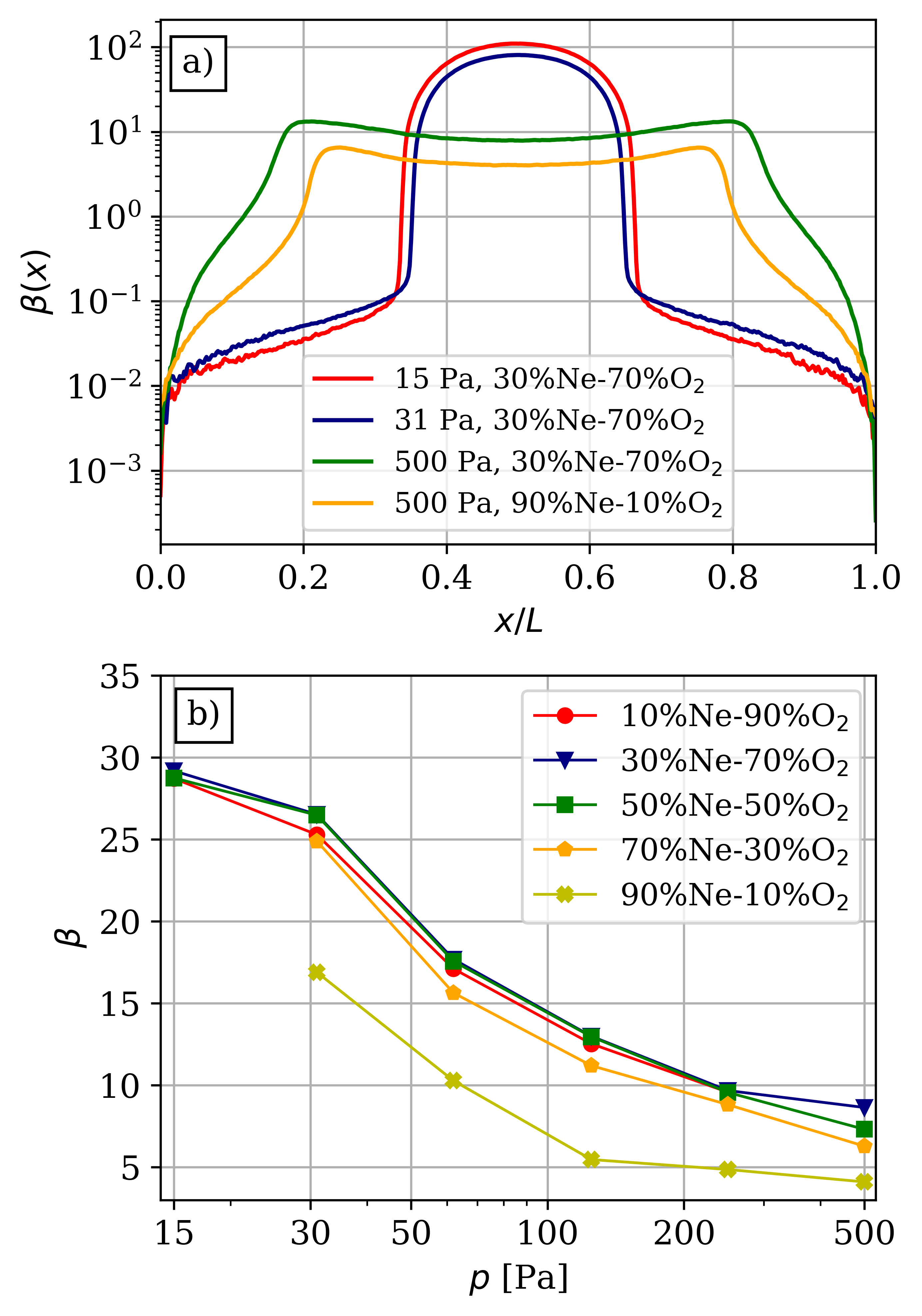}
	\caption{(a) Time averaged ratio of the negative ion (O$^-$) density and electron density (\textit{local} electronegativity, $\beta(x)$) for different pressures and neon/oxygen concentrations of the background gas mixture (for the same discharge conditions as those in figure~\ref{fig:dens}).
	(b) \textit{Global} electronegativity of the discharge, $\beta$ as a function of pressure for different neon/oxygen concentrations of the background gas mixture. Discharge conditions: $L=2.5$~cm, $f=10$~MHz, $V_{\rm{pp}} = 350$~V.}
	\label{fig:electronegativity}
\end{figure}

\begin{figure}[ht]
	\centering
	\includegraphics[width=0.5\linewidth]{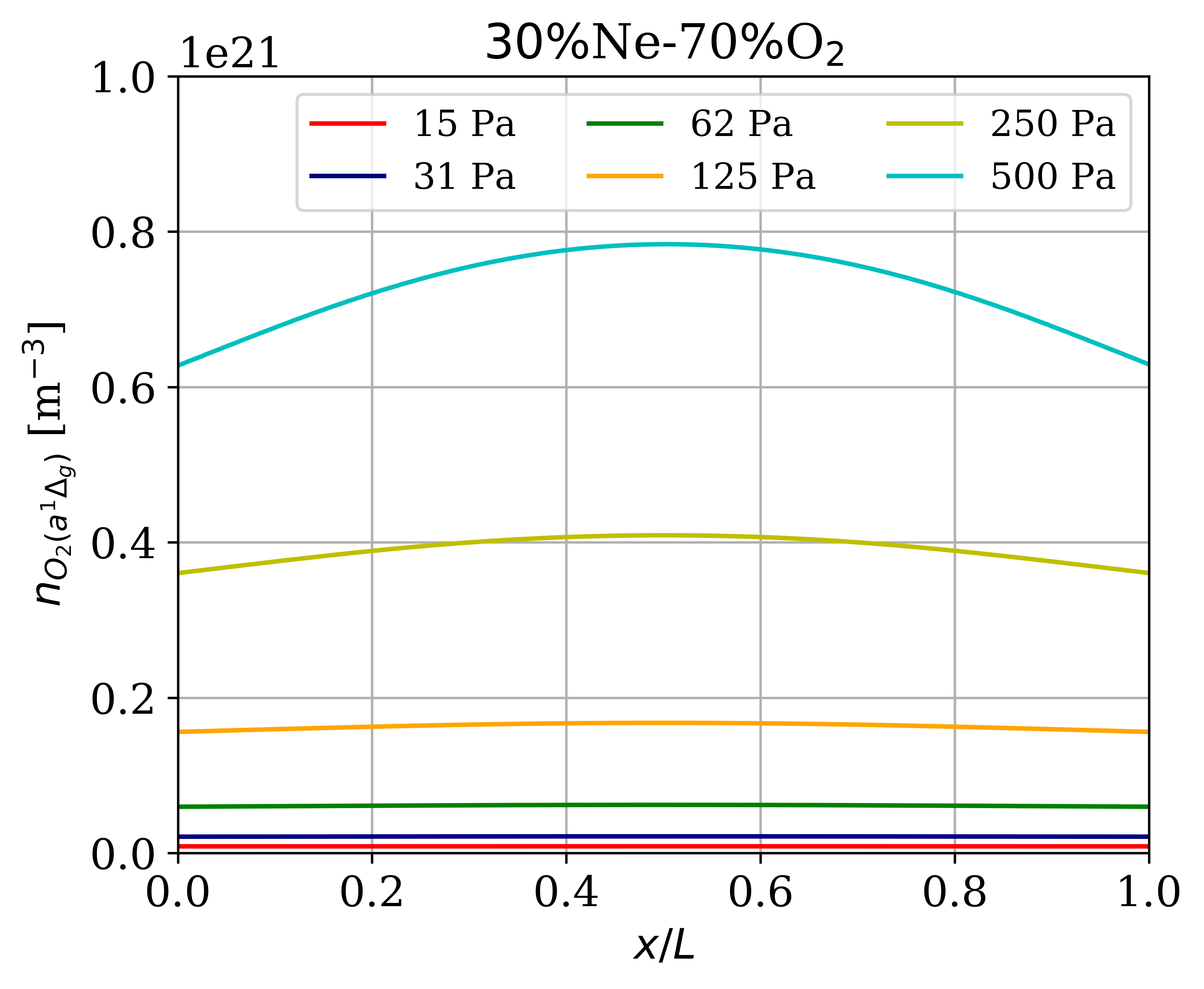}
	\caption{Time averaged $\rm{O_2 (a^1 \Delta_g)}$ singlet delta oxygen metastable molecule concentration obtained for 30\%~Ne--70\%~O$_2$ gas mixture at different pressures. Discharge conditions: $L=2.5$~cm, $f=10$~MHz, $V_{\rm{pp}} = 350$~V.}
	\label{fig:meta}
\end{figure}

Figure~\ref{fig:dens} shows PIC/MCC simulation results on the time averaged density distributions of charged particles and fast neutrals traced in the simulations for exemplary cases from the wide parameter regime. Panels of figure~\ref{fig:dens} show results obtained for different neutral gas pressures and different neon/oxygen concentration of the background gas mixture: (a) 15~Pa, 30\%~Ne--70\%~O$_2$, (b) 31~Pa, 30\%~Ne--70\%~O$_2$, (c) 500~Pa, 30\%~Ne--70\%~O$_2$, and (d) 500~Pa, 90\%~Ne--10\%~O$_2$. (Note that the PROES results corresponding to these cases are shown in panels (b1), (b2), (b6), and (e6) of figure~\ref{fig:exp}, respectively.)
At 15~Pa, 30\%~Ne--70\%~O$_2$ mixture (panel (a)), the time averaged electron density exhibits local maxima at the edges of the narrow bulk region, and it is significantly lower than the density of negative O$^-$ ions in the bulk. The electronegativity is very high in the discharge center, the ratio of the negative ion density and electron density is about 100 (figure~\ref{fig:electronegativity}(a)). The global electronegativity (the ratio of the density of negative ions and electrons averaged over the electrode gap) is about 30 under these conditions (see figure~\ref{fig:electronegativity}(b)). The density of Ne$^+$ ions is about three orders of magnitude lower than the density of O$_2^+$ ions in the discharge center and the density of fast neutral species is high over the whole discharge gap.
In this gas mixture, local maxima in the time averaged electron density distribution at the edges of the bulk region and the high negative ion density in the bulk are found also at a higher pressure of 31~Pa (panel (b)). However, at 31~Pa the electron density is enhanced in the discharge center compared to the 15~Pa case, while the density of Ne$^+$ ions, as well as the density of fast neutrals, decreases in the bulk. The electronegativity is maximum in the discharge center (figure~\ref{fig:electronegativity}(a)), the global electronegativity of the discharge is about 25 (figure~\ref{fig:electronegativity}(b)).
By increasing the pressure to 500~Pa for the 30\%~Ne--70\%~O$_2$ mixture (panel (c)), the time averaged electron density exhibits its maximum in the discharge center and local minima at the bulk edges. The density of Ne$^+$ ions as well as the density of fast neutrals drops in the bulk. The bulk region is wide at this high pressure. The O$^-$ density profile has a local minimum in the discharge center and local maxima near the edges of the bulk region. Consequently, the electronegativity has a local minimum in the center and local maxima at the bulk-sheath boundary (figure~\ref{fig:electronegativity}(a)). In the discharge center, the density of negative ions is about one order of magnitude higher than the electron density. The global electronegativity is about  8 under these conditions (figure~\ref{fig:electronegativity}(b)).
By decreasing the O$_2$ concentration in the mixture to 10\% at this high pressure (figure~\ref{fig:dens}(d)), the density of electrons increases, while the O$^-$ density remains high in the bulk region. The O$^-$ density is about four times higher than the electron density in the discharge center. The global electronegativity has a low value of about 2 under these conditions, i.e. the discharge remains electronegative also in case of the lowest O$_2$ concentration of 10\% in the mixture. As it can be seen in figure~\ref{fig:electronegativity}(b), the global electronegativity decreases with increasing the pressure in all mixtures. For O$_2$ concentrations above 30\% the electronegativity does not depend on the Ne/O$_2$ concentration ratio in the background gas mixture.  

The simulation results show that the self-consistently calculated density of singlet delta oxygen metastable molecules increases with the pressure. However, the $\rm{O_2 (a^1 \Delta_g)}$ density remains low in all cases, below 3\% of the density of background O$_2$ molecules. Figure~\ref{fig:meta} shows the time averaged density distribution of $\rm{O_2 (a^1 \Delta_g)}$ molecules in the gap as a function of pressure for 30\%~Ne--70\%~O$_2$ gas mixture. At low pressures, the metastable density profiles are flat, while at high pressures, profiles with peak metastable density in the center of the discharge are obtained. This is due to the pressure dependence of the diffusion coefficient (see eq.~\ref{eqn:diff}, at high pressure the diffusion slows down).

\begin{figure}[ht]
	\centering
	\includegraphics[width=0.5\linewidth]{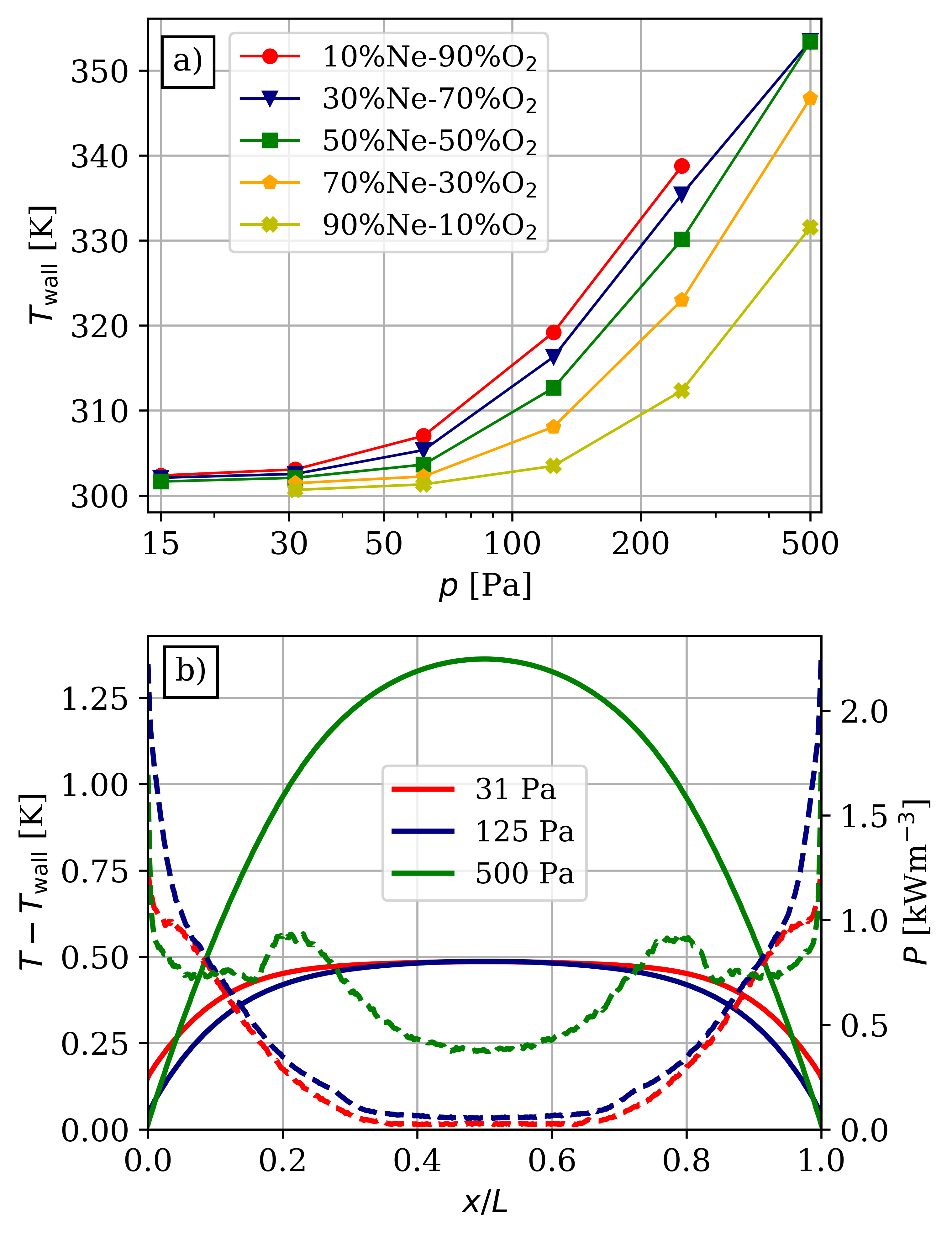}
	\caption{(a) Wall temperature, $T_{\rm wall}$ as a function of pressure for different Ne-O$_2$ concentrations of the background gas mixture. (b) Spatial distribution of the gas temperature with respect to the wall temperature (solid lines; left vertical axis), and spatial distribution of the thermal power input density, $P(x)$ (dashed lines; right vertical axis) for three different pressure values in case of 30\%~Ne--70\%~O$_2$ gas mixture. Discharge conditions: $L=2.5$~cm, $f=10$~MHz, $V_{\rm{pp}} = 350$~V.
	}
	\label{fig:T}
\end{figure}

The change of the wall temperature as a function of pressure for various Ne/O$_2$ concentrations of the background gas mixture is shown in figure~\ref{fig:T}(a). The wall temperature, $T_{\rm wall}$, increases significantly compared to the initial wall temperature (which was set to 300~K in the simulations) with increasing the pressure. At a given pressure, $T_{\rm wall}$ decreases with decreasing the O$_2$ content in the gas mixture. This is due to the different efficiency of the power absorption of Ne and O$_2$ at identical discharge conditions.
In panel (b) of  figure~\ref{fig:T} the spatial distribution of the gas temperature with respect to the wall temperature (left vertical axis)
and the power input field (right vertical axis) are shown at three different pressures for the 30\%~Ne--70\%~O$_2$ case.  The temperature profiles show the maximum gas temperature in the discharge center at all pressures, which is only slightly higher than the temperature of the electrodes. Compared to the wall temperature, an increase by only about 1.5~K in the discharge center is found at the highest pressure. The simulations revealed that at the low electrical power levels of the present setup, being only a few Watts, the increased temperature of the gas is mainly due to the increased temperature of the electrodes, and the gas temperature does not significantly change within the gap. 

\begin{figure}[ht]
	\centering
	\includegraphics[width=0.95\linewidth]{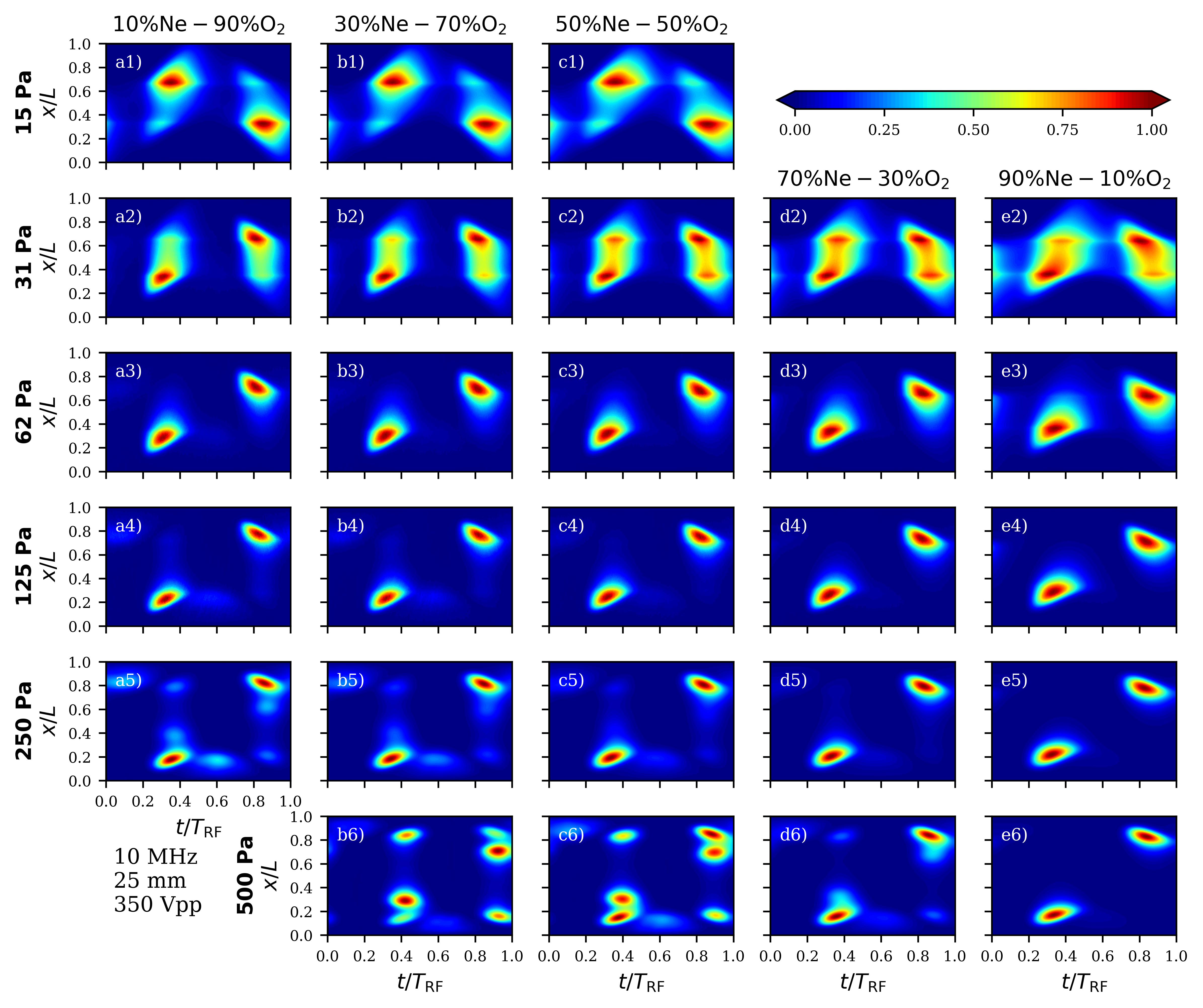}
	\caption{Spatio-temporal plots of the electron-impact excitation rate
	[a.u.] from the ground state into the Ne~$\rm{3p_0}$ state obtained from PIC/MCC simulations of in Ne-O$_2$ CCPs at different neutral gas pressures: 15~Pa (first row), 31 Pa (second row), 62~Pa (third row), 125~Pa (fourth row), 250~Pa (fifth row) and 500~Pa (sixth row), for different neon/oxygen concentration of the background gas mixture: 10\%~Ne--90\%~O$_2$ (first column), 30\%~Ne--70\%~O$_2$ (second column), 50\%~Ne--50\%~O$_2$ (third column), 70\%~Ne--30\%~O$_2$ (fourth column), 90\%~Ne--10\%~O$_2$ (fifth column). 
	The powered electrode is located at $x/L=0$, while the grounded electrode is at $x/L=1$. Discharge conditions: $L=2.5$~cm, $f=10$~MHz, $V_{\rm{pp}} = 350$~V.}
	\label{fig:sim1}
\end{figure}

Figure~\ref{fig:sim1} shows the spatio-temporal distribution of the electron impact excitation rate from the ground state into the Ne~$\rm{3p_0}$ state obtained from PIC/MCC simulations for discharge conditions identical to those presented in Figure~\ref{fig:exp}. Similarly to Figure~\ref{fig:exp}, panels in a given row show the results obtained for a given pressure, for different Ne--O$_2$ mixtures, for decreasing O$_2$ concentration in panels from left to right.  
At the lowest pressure of 15~Pa (first row), the spatio-temporal distribution of the electron impact excitation rate measured by PROES is well reproduced by the simulations for all mixtures, the excitation patterns I.--III. observed by PROES at this pressure can be identified in the simulation results as well. At 31~Pa (second row), in agreement with the PROES results, the strongest excitation is found at the expanding sheath edges (pattern I.) in all mixtures.
At 62~Pa and 125~Pa (third and fourth rows), the simulation results are in good agreement with the PROES results in all cases.
At 250~Pa (fifth row), the formation of the two distinct excitation patterns at the edges of the bulk region (IV. and V.) observed in the PROES results are also captured in the simulations, as well as the presence of the additional excitation pattern (VI.) at the phase of maximum sheath expansion at both electrodes.
At the highest pressure of 500~Pa (sixth row), the main excitation patterns revealed by PROES are  reproduced by the simulations (including their relative intensity). 

In the following, the simulation results are analysed in detail in four representative cases: (i) 15~Pa, 30\%~Ne--70\%~O$_2$, (ii) 31~Pa, 30\%~Ne--70\%~O$_2$, (iii) 500~Pa, 30\%~Ne--70\%~O$_2$, and (iv) 500~Pa, 90\%~Ne--10\%~O$_2$. For these cases, the spatio-temporal plots of the electron-impact excitation rate from the ground state into the Ne~$\rm{3p_0}$ state are shown in panels (b1), (b2), (b6), and (e6) of figure~\ref{fig:exp} (PROES) and figure~\ref{fig:sim1} (PIC/MCC simulation), respectively. Under these conditions, the main excitation patterns identified above (labeled I.-VI. in figure~\ref{fig:exp}) are exhibited. The time averaged particle density distributions and the electronegativity obtained for these four cases were presented in figure~\ref{fig:dens} and figure~\ref{fig:electronegativity}(a), respectively, and discussed above. 

\begin{figure}[ht]
	\centering
	\includegraphics[width=0.95\linewidth]{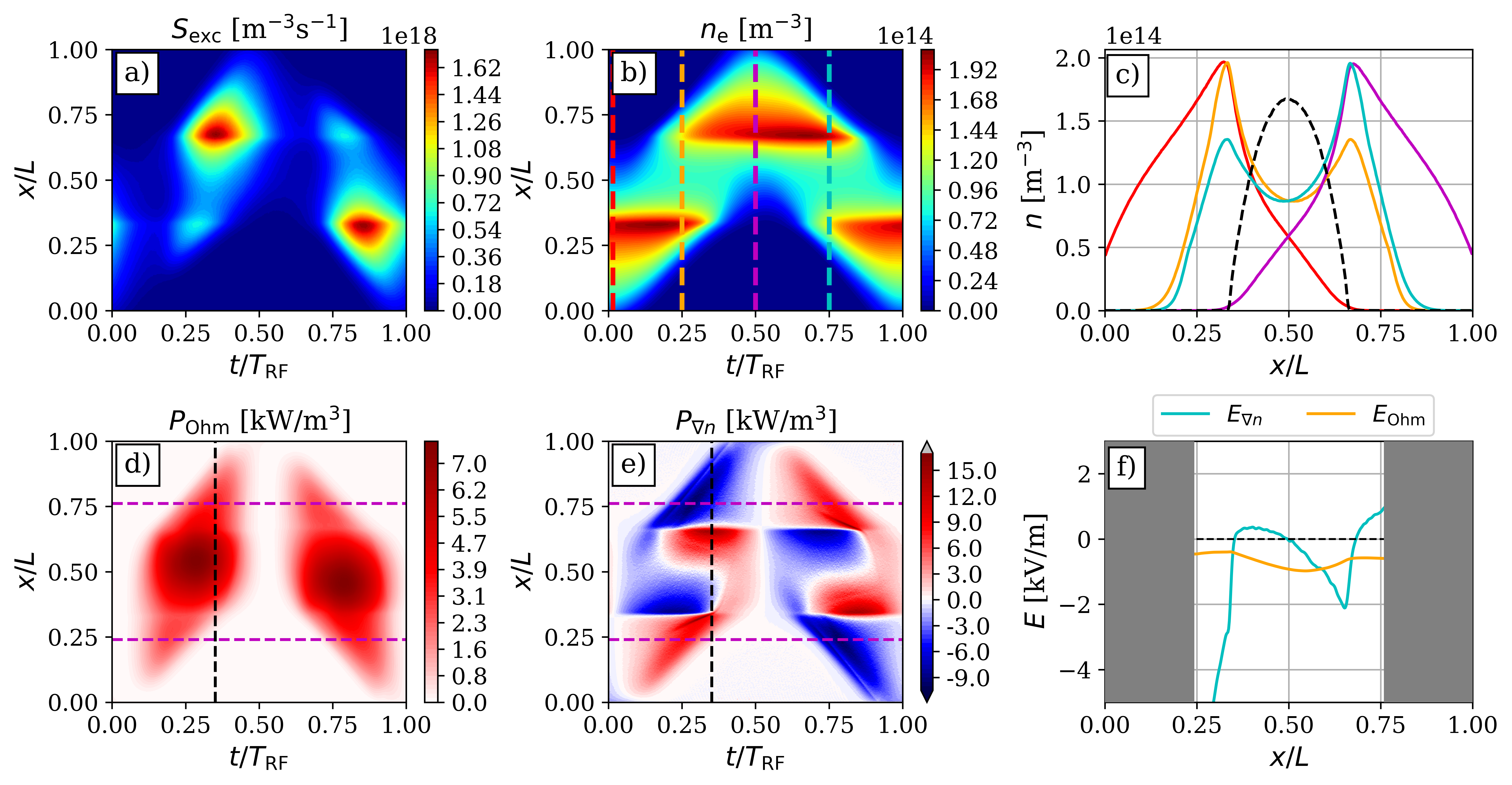}
	\caption{Spatio-temporal plot of the electron-impact excitation rate, $S_{\rm exc}$ (a), the electron density, $n_{\rm e}$ (b), temporal snapshots of the electron density (solid color curves) as well as the temporally averaged negative ion (O$^-$) density (dashed black curve) divided by a factor of 50 (c), spatio-temporal plot of the Ohmic power absorption, $P_{\rm Ohm}$ (d), of the ambipolar power absorption, $P_{\nabla n}$ (d) as well as temporal snapshots of the Ohmic and the ambipolar electric field, $E_{\rm Ohm}$ and $E_{\nabla n}$, respectively (f) for 15 Pa and 30\%~Ne--70\%~O$_2$ gas mixture. The color of the curves in panel (c) denote the respective time instance in panel (b). The vertical dashed black lines in panels (d,e) indicate the time instance at which panel (f) is plotted. The horizontal dashed magenta lines indicate the edges of the masked regions in panel (f). 	The powered electrode is located at $x/L=0$, while the grounded electrode is at $x/L=1$. Discharge conditions: $L=2.5$~cm, $f=10$~MHz, $V_{\rm{pp}} = 350$~V.} 
	\label{fig:15Pa}
\end{figure}

Figure~\ref{fig:15Pa} shows PIC/MCC simulation results for various discharge characteristics obtained for 30\%~Ne--70\%~O$_2$ gas mixture at 15 Pa. The spatio-temporal distribution of the electron-impact excitation rate from the ground state into the Ne~${\rm 3p_0}$ state, $S_{\rm exc}$ is shown in panel (a) (not normalized, otherwise same as panel (b1) of figure~\ref{fig:sim1}), while that of the electron density, $n_{\rm e}$ is shown in panel (b). In panel (b), four time instances within the RF period are marked with vertical dashed lines in different colours. The electron densities in the discharge gap corresponding to these time instances, i.e. temporal snapshots of the electron density are shown in panel (c), where the color of the solid curves denote the respective time instance in panel (b). At all selected time instances, sharp electron density peaks can be seen at the edges of the bulk region (one peak at the powered/grounded electrode side at $t/T_{\rm RF}$ values of 0 and 0.5; peaks at both sides at $t/T_{\rm RF}$ values of 0.25 and 0.75), while the electron density is low in the discharge center. In panel (c) the temporally averaged negative ion (O$^-$) density divided by a factor of 50 is also included (dashed black line) to illustrate that the density of O$^-$ ions is much higher than the electron density in the bulk region. As it can be seen in figure~\ref{fig:dens}(a), the time averaged O$^-$ density is about two orders of magnitude higher than the time averaged electron density in the discharge center. As a consequence of this, the ratio of the O$^-$ ion and electron density, i.e. the electronegativity is high (about 100) in the discharge center (figure~\ref{fig:electronegativity}(a)). Under these conditions the ratio of negative ions and electrons averaged over the electrode gap, i.e. the global electronegativity is also high (about 30, see figure~\ref{fig:electronegativity}(b)).    
The spatio-temporal plots of the Ohmic power absorption, $P_{\rm Ohm}$, and the ambipolar power absorption, $P_{\nabla n}$, are shown in panels (d) and (e), respectively. 
Due to the high electronegativity in the discharge center, the conductivity of the plasma is low in the bulk region. Consequently, the Ohmic power absorption, $P_{\rm Ohm}$, peaks in the discharge center at the times of maximum RF current within the RF period, as it can be seen in panel~(d). The ambipolar power absorption, $P_{\nabla n}$, peaks at the edges of the bulk region at the time of sheath collapse at both electrodes (see panel (e)) due to the local maximum of the electron density and the corresponding large gradients in the electron density (panels (b) and (c)). Under these conditions, the ambipolar power absorption is the dominant power absorption mechanism.
Panel (f) shows temporal snapshots of the Ohmic and ambipolar electric field, $E_{\rm Ohm}$ and $E_{ \nabla p}$, respectively, at a selected time instance indicated by black vertical dashed lines in panels (d) and (e). At this time, the Ohmic electric field has a negative sign in the bulk, while the ambipolar electric field changes the sign three times. The sum of these two terms results in a strong electric field at the grounded electrode side of the bulk (at the instantaneous collapsing sheath side) and a weak, almost zero electric field at the powered electrode side of the bulk (at the instantaneous expanding sheath side). 
Electrons accelerated by these electric fields induce strong excitation (III.) at the collapsing sheath edge and weaker excitation (II.) in the discharge center (see panel (a)). Weak excitation at the expanding sheath edge (I.) can be also observed. These are excitation patterns characteristic of a hybrid $\alpha$-DA discharge operation mode with dominant DA-mode, specific to electronegative discharges (panel (a)). 
\begin{figure}[ht]
	\centering
	\includegraphics[width=0.95\linewidth]{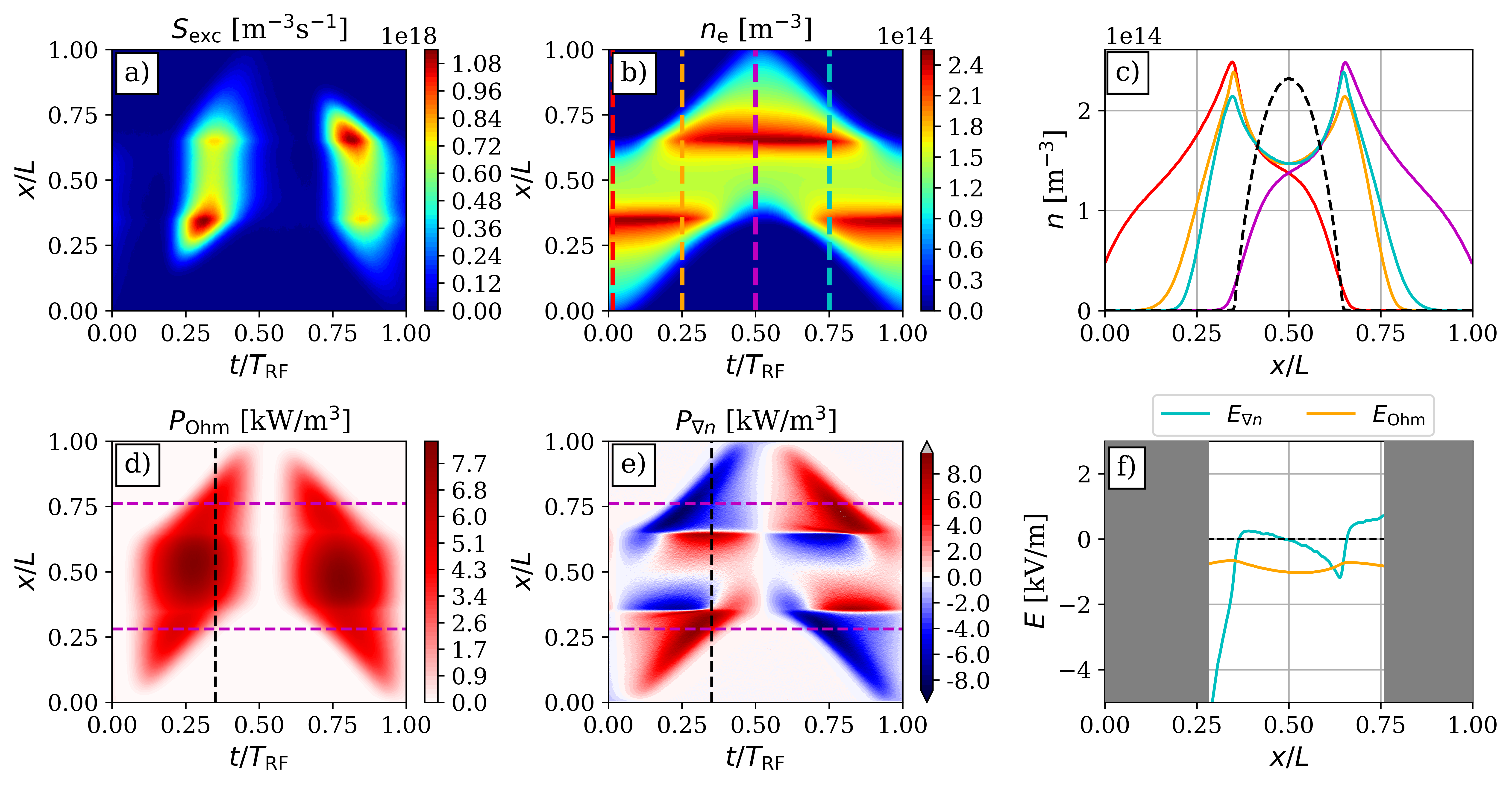}
	\caption{Various discharge characteristics -- same as those presented in figure~\ref{fig:15Pa} -- obtained for 31~Pa and 30\%~Ne--70\%~O$_2$ gas mixture. Discharge conditions: $L=2.5$~cm, $f=10$~MHz, $V_{\rm{pp}} = 350$~V.}
	\label{fig:31Pa}
\end{figure}

Discharge characteristics obtained from the simulations performed for the same gas mixture as in case of Figure~\ref{fig:15Pa}, i.e. 30\%~Ne--70\%~O$_2$, but at a higher pressure of 31~Pa, are shown in Figure~\ref{fig:31Pa}. In this case, the spatio-temporal distribution of the excitation rate is dominated by the $\alpha$-peaks (I.) in combination with drift features (II.) in the bulk and ambipolar peaks (III.) at the collapsing sheath edges. Similarly to the 15~Pa case, the electron density profiles plotted at different time instances show strong peaks at the edges of the bulk region (panel (c)). The electron
density is low in the bulk (panels (b) and (c)), however it is enhanced in the discharge center compared to the 15~Pa case. Similarly to the 15~Pa case, the electronegativity is maximum in the discharge center (figure~\ref{fig:electronegativity}(a)) and the global electronegativity has a high value of about 25 (figure~\ref{fig:electronegativity}(b)).
The Ohmic power absorption peaks in the discharge center (panel (d)) and its magnitude is similar to that obtained at 15~Pa. The ambipolar power absorption peaks at the edges of the bulk region (panel (e)) and its magnitude is significantly reduced compared to the 15~Pa case. The superposition of the Ohmic electric field and the ambipolar electric field at the selected time instance (panel (f)) results in a strong electric field at the powered electrode side of the bulk (at the instantaneous expanding sheath side) as well as at the grounded electrode side of the bulk (at the instantaneous collapsing sheath side). The electrons accelerated by these electric fields induce the strong excitation peak (I.) at the expanding sheath edge and the weaker excitation peak (III.) at the collapsing sheath edge. The nonzero electric field in the discharge center leads to excitation in the central bulk region (feature II.) These are excitation patterns characteristic of a hybrid $\alpha$-DA discharge operation mode with dominant $\alpha$-mode (panel (a)). 

\begin{figure}[ht]
	\centering
	\includegraphics[width=0.95\linewidth]{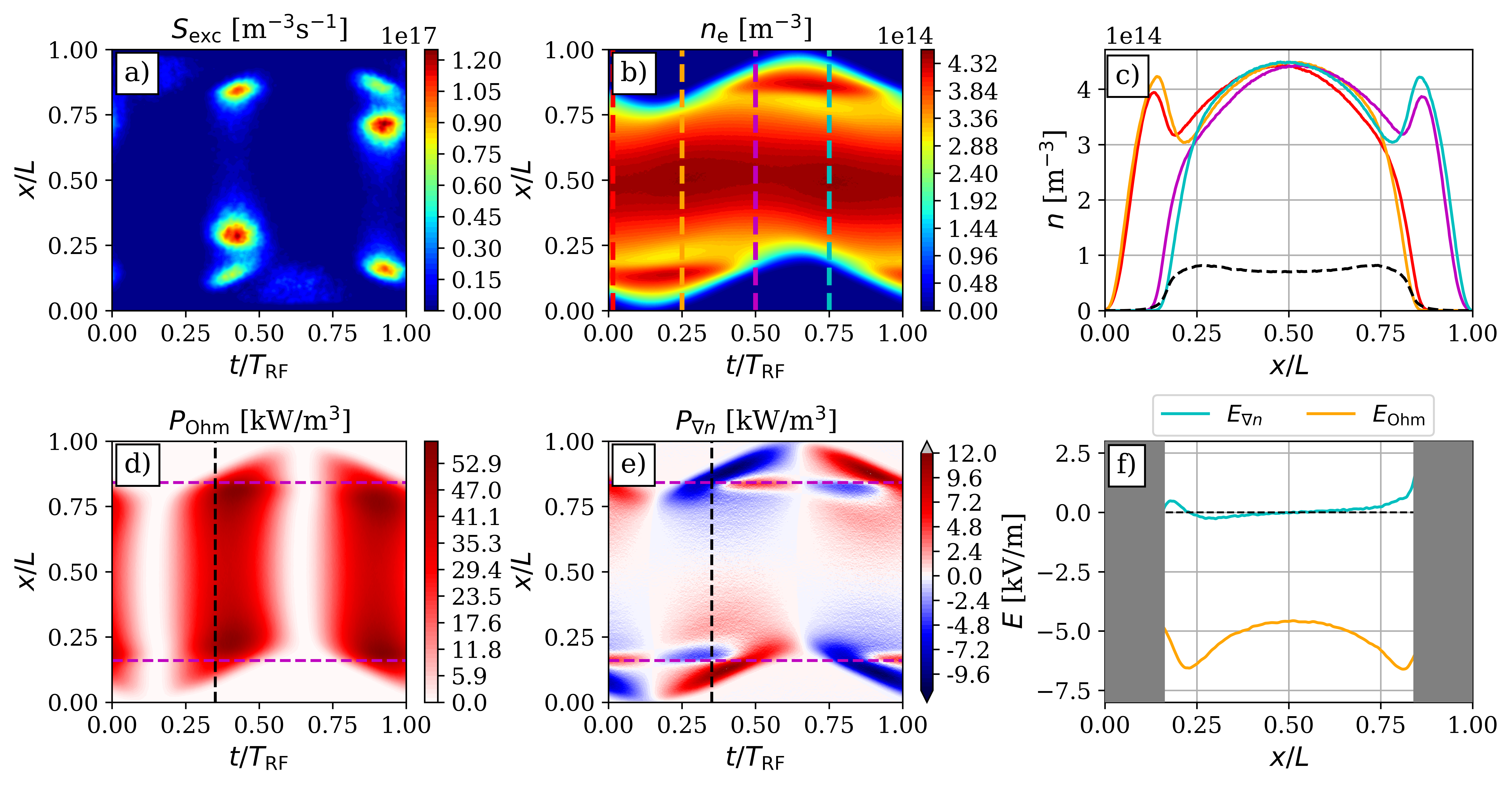}
	\caption{
	Various discharge characteristics -- same as those presented in figure~\ref{fig:15Pa} -- obtained for 500~Pa and 30\%~Ne--70\%~O$_2$ gas mixture. Discharge conditions: $L=2.5$~cm, $f=10$~MHz, $V_{\rm{pp}} = 350$~V.}
	\label{fig:500PaNe30}
\end{figure}

The discharge characteristics obtained at a very high pressure of 500~Pa in 30\%~Ne--70\%~O$_2$ gas mixture are shown in figure~\ref{fig:500PaNe30}. Under these conditions a weak $\alpha$-peak (I.) and two additional excitation peaks (IV. and V.) in about the same time interval are found in both halves of the RF period, as well as weak excitation patterns (VI.) in the sheaths (panel (a)). The electron density distributions at different time instances show a local minimum close to the  bulk edges, at about the position of maximum sheath expansion at both electrodes and maximum in the center of the discharge (panels (b) and (c)). The time averaged density of negative ions has a local minimum in the center and local maxima at the borders of the sheath region (note that the time averaged O$^-$ density is divided by 50 in panel (c)). 
The reason for this is the increasing locality of the charged particle transport as a function of pressure. During sheath expansion, energetic electrons are created at the expanding sheath edge, which induce a local formation of O$^-$ ions via high-energy threshold dissociative attachment (reaction 20 in table~\ref{table:Neon-Oxygen-mixture}). At low pressures, these processes are not localized at the sheath edge.
As a consequence of these characteristics, the electronegativity has a local maximum at about the positions of local minima in the electron density (figure~\ref{fig:electronegativity}(a)), which leads to the development of drift electric fields in these regions. These are enhanced by the bulk electric field induced by the high collisionality of the plasma at this high pressure. Therefore, both the collisionality due to the high pressure and the electronegativity of the discharge contribute to the ohmic power absorption term. As a result of the superposition of these two effects, the Ohmic power absorption peaks at the edges of the bulk (panel (d)), in the same regions where the excitation patterns IV. and V. are found. Compared to the Ohmic power absorption, the ambipolar power absorption (panel (e)) is less significant under these conditions. This can be seen also in panel (f) which shows the Ohmic and ambipolar electric fields at the selected time instance: the ambipolar electric field is weak compared to the Ohmic field. The Ohmic field plotted at the time instance indicated in panel (d) shows local maxima at the bulk edges which are responsible for the 
excitation peaks IV. and V. at the bulk edges. 

\begin{figure}[ht]
	\centering
	\includegraphics[width=0.95\linewidth]{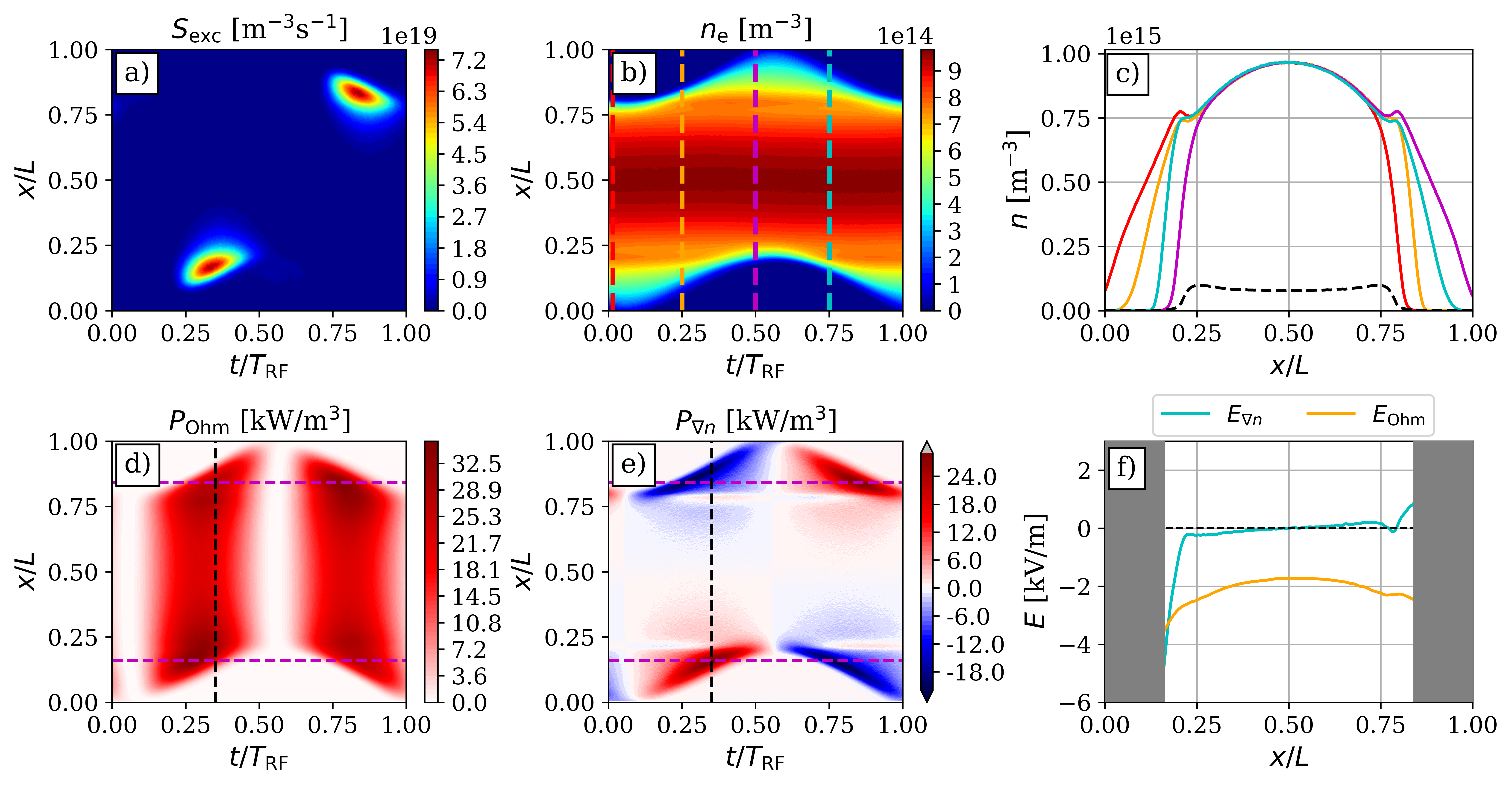}
	\caption{
	Various discharge characteristics -- same as those presented in figure~\ref{fig:15Pa} -- obtained for 500~Pa and 90\%~Ne--10\%~O$_2$ gas mixture. Discharge conditions: $L=2.5$~cm, $f=10$~MHz, $V_{\rm{pp}} = 350$~V.}
	\label{fig:500PaNe90}
\end{figure}

\begin{figure}[ht]
	\centering
	\includegraphics[width=0.95\linewidth]{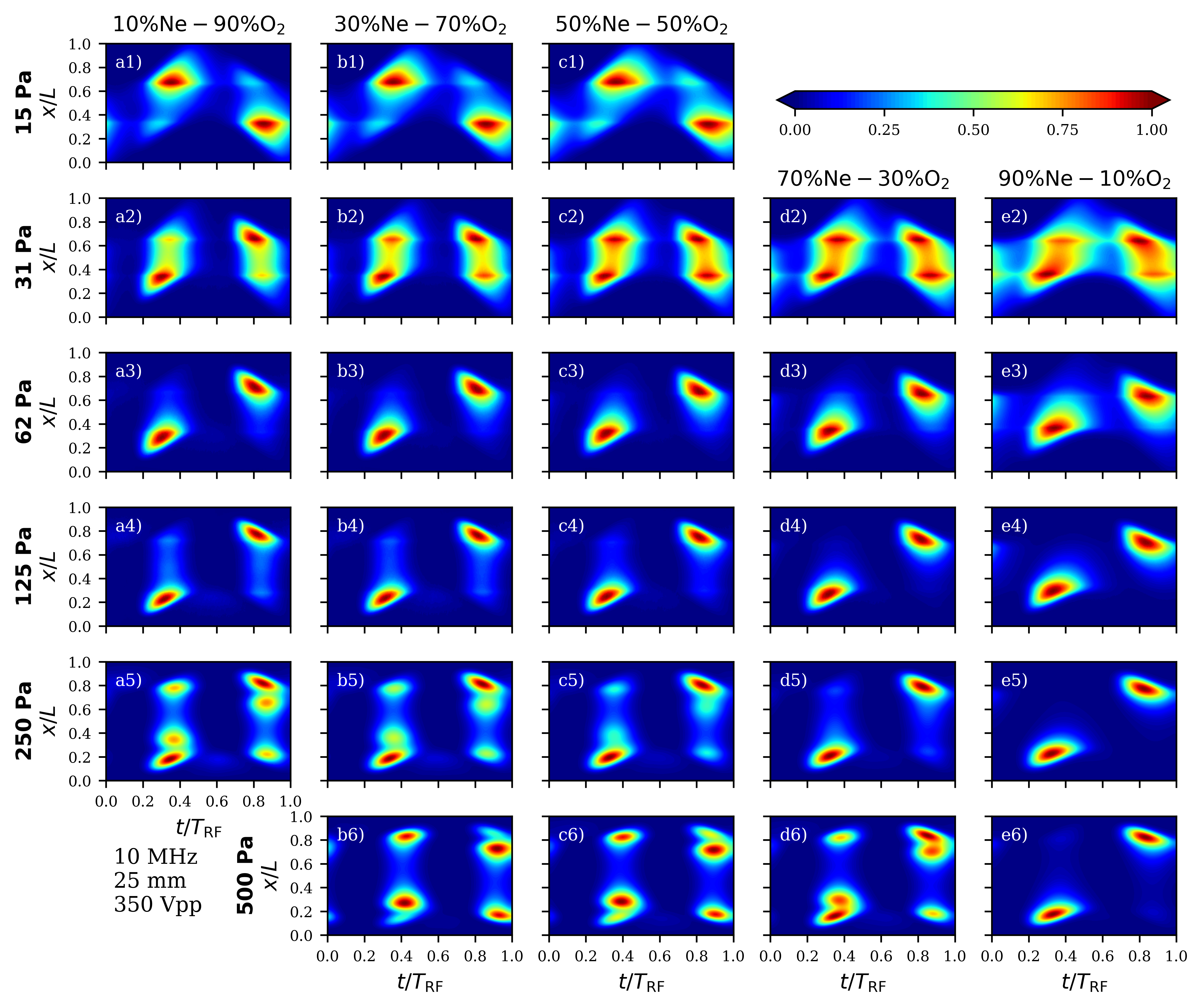}
	\caption{Spatio-temporal plots of the oxygen ionization rate 
	[a.u.] obtained from PIC/MCC simulations of Ne-O$_2$ CCPs at different neutral gas pressures: 15~Pa (first row), 31 Pa (second row), 62~Pa (third row), 125~Pa (fourth row), 250~Pa (fifth row) and 500~Pa (sixth row), for different neon/oxygen concentration of the background gas mixture: 10\%~Ne--90\%~O$_2$ (first column), 30\%~Ne--70\%~O$_2$ (second column), 50\%~Ne--50\%~O$_2$ (third column), 70\%~Ne--30\%~O$_2$ (fourth column), 90\%~Ne--10\%~O$_2$ (fifth column). 
	The powered electrode is located at $x/L=0$, while the grounded electrode is at $x/L=1$. Discharge conditions: $L=2.5$~cm, $f=10$~MHz, $V_{\rm{pp}} = 350$~V.}
	\label{fig:sim2}
\end{figure}

The simulation results obtained at the highest pressure of 500~Pa by increasing the Ne concentration to 90\%, i.e. for the  90\%~Ne--10\%~O$_2$ gas mixture, are shown in figure~\ref{fig:500PaNe90}. Under these conditions strong excitation is found only at the expanding sheath edges (pattern I.) at both electrodes (panel (a)). The electron density profiles  for the different time instances show only small local minima at the edges of the bulk region (panel~(c)). The time averaged O$^-$ density has a local minimum in the center and local maxima at the borders of the sheath region. The discharge remains electronegative (the global electronegativity is about 2, see figure~\ref{fig:electronegativity}(b)). The spatio-temporal distribution of the Ohmic power absorption (panel~(d)) shows peaks at the edges of the bulk, but these are less pronounced compared to the case of high O$_2$ concentration at the same pressure. The ambipolar power absorption (panel (e)) is mainly concentrated at the expanding sheath edges and has about the same magnitude as the Ohmic power absorption. At this low O$_2$ concentration of 10\% in the mixture, there are no local maxima in the Ohmic electric field at the bulk edges at the selected time instance, and the ambipolar electric field exhibits only slight modulation in the gap (panel (f)). Under these conditions, the discharge operates in $\alpha$-mode. 

In summary, based on the PIC/MCC simulation results, the mechanisms behind the development of the various excitation features revealed by the PROES measurements in Ne-O$_2$ CCPs in the wide pressure range for different gas mixing ratios have been successfully revealed.

Finally, figure~\ref{fig:sim2} shows the spatio-temporal distribution of the electron impact ionization rate of oxygen obtained from PIC/MCC simulations for different pressures and different Ne-O$_2$ mixtures (for the same conditions as those covered in case of the PROES results presented in figure~\ref{fig:exp} and the excitation rates in figure~\ref{fig:sim1}). The ionization patterns seen in panels of figure~\ref{fig:sim2} reflect the excitation patterns of figure~\ref{fig:sim1}. This means that the PROES measurement results in mixtures of Ne and O$_2$ under the conditions studied probe both the excitation and the ionization dynamics in the discharge. Under these conditions, the PROES measurements provide correct information on the discharge operation mode. 

\section{Conclusions}\label{sec:Conclusions}

We have performed Phase Resolved Optical Emission Spectroscopy (PROES) measurements combined with 1d3v Particle-in-Cell/Monte Carlo Collisions (PIC/MCC) simulations in low-pressure capacitively coupled neon-oxygen gas mixture plasmas. This study covered a wide pressure range and a wide mixing range of Ne and O$_2$ gases for a geometrically symmetric plasma reactor with a gap length of 2.5~cm, operated at a driving frequency of 10~MHz and a peak-to-peak voltage of 350~V. The pressure of the gas mixture was varied between 15~Pa and 500~Pa, while the neon/oxygen concentration was changed between 10\% and 90\%. 

For all discharge conditions, the spatio-temporal distribution of the electron-impact excitation rate from the Ne ground state into the Ne~$\rm{3p_0}$ state was recorded by PROES. The measured electron-impact excitation rate from the Ne ground state into the Ne~$\rm{3p_0}$ state was compared to the PIC/MCC simulation results on the Ne excitation rate, resulting in a good qualitative agreement in the whole parameter regime. 
At the lowest pressure, a weak $\alpha$-peak at the expanding sheath edge, strong ambipolar peak, and weak drift feature in the bulk region were found in the excitation rates. With increasing pressure, the $\alpha$-peak and the drift feature were found to be enhanced, while the ambipolar peak was reduced. At intermediate pressures, the $\alpha$-peak was found to be the dominant excitation pattern in all mixtures. Further increase of the pressure resulted in the formation of two distinct excitation peaks at the edges of the bulk region, which dominated the excitation at high O$_2$ concentrations.

Based on the emission/excitation patterns, multiple discharge operation regimes were identified. It was found that the localized bright emission features at the bulk boundaries at high pressures and high O$_2$ concentrations are caused by local maxima in the electronegativity. The relative contributions of the ambipolar and the Ohmic electron power absorption were found to vary strongly with the discharge parameters: the Ohmic power absorption was enhanced by both the high collisionality at high pressures and the high electronegativity at low pressures. In the wide parameter regime covered in this study, the PROES measurements were found to accurately probe the ionization dynamics in the discharge, i.e. the discharge operation mode.

The simulation revealed that the temperature of the electrodes increases significantly compared to the initial wall temperature with increasing the gas pressure. It was found that the power deposition within the gas causes only a slight increase of the gas temperature above the temperature of the electrodes, which was, however, found to be significant due to the heating of the electrodes by the particles from the plasma. This finding points out the importance of the thermal balance of the electrode construction in determining the electrode and gas temperatures under operating conditions at moderate electrical power levels.

\ack This work was supported by the Hungarian National Research, Development and Innovation Office via grants K-134462 and FK-128924, by the German Research Foundation (DFG) within the frame of the collaborative research centre SFB-TR 87 (project C1), SFB-CRC 1316 (project A4), by the project, ``Electron heating in capacitive RF plasmas based on moments of the Boltzmann equation: from fundamental understanding to knowledge based process control'' (No. 428942393), and by the \'{U}NKP-21-1 New National Excellence Program of the Ministry for Innovation and Technology from the source of the National Research, Development and Innovation Fund.

\section*{References}
	

\begin{thebibliography}{10}
\expandafter\ifx\csname url\endcsname\relax
  \def\url#1{{\tt #1}}\fi
\expandafter\ifx\csname urlprefix\endcsname\relax\def\urlprefix{URL }\fi
\providecommand{\eprint}[2][]{\url{#2}}

\bibitem{Liebermann_book}
Lieberman M~A and Lichtenberg A~J 2005 {\em Principles of Plasma Discharges and
  Materials Processing, 2nd Edition\/} (Wiley)

\bibitem{Makabe_book}
 T. Makabe and Z. Petrovi\'c, {\em Plasma Electronics: Applications in
  Microelectronic Device Fabrication} (Taylor \& Francis, London, 2006).

\bibitem{Chabert_book}
Chabert P and Braithwaite N 2011 {\em Physics of Radio-Frequency Plasmas\/}
  (New York: Cambridge University Press)

\bibitem{Makabe08}
Ohmori T and Makabe T 2008 {\em Applied Surface Science\/} {\bf 254} 3696--3709

\bibitem{Makabe_2019}
Makabe T 2019 {\em Japanese Journal of Applied Physics\/} {\bf 58} 110101

\bibitem{Belenguer1990}
Belenguer P and Boeuf J~P 1990 {\em Physical Review A\/} {\bf 41} 4447--4459

\bibitem{Schulze2011}
Schulze J, Derzsi A, Dittmann K, Hemke T, Meichsner J and Donk\'o Z 2011 {\em
  Physical Review Letters\/} {\bf 107} 275001

\bibitem{Liu2016}
Liu Y~X, Sch\"ungel E, Korolov I, Donk\'o Z, Wang Y~N and Schulze J 2016 {\em
  Physical Review Letters\/} {\bf 116} 255002

\bibitem{Liu2017}
Liu Y~X, Korolov I, Schüngel E, Wang Y~N, Donk{\'{o}} Z and Schulze J 2017
  {\em Plasma Sources Science and Technology\/} {\bf 26} 055024

\bibitem{Wang_2019}
Wang L, Wen D~Q, Zhang Q~Z, Song Y~H, Zhang Y~R and Wang Y~N 2019 {\em Plasma
  Sources Science and Technology\/} {\bf 28} 055007

\bibitem{Skarphedinsson_2020}
Skarphedinsson G~A and Gudmundsson J~T 2020 {\em Plasma Sources Science and
  Technology\/} {\bf 29} 084004

\bibitem{Proto_2021}
Proto A and Gudmundsson J~T 2021 {\em Plasma Sources Science and Technology\/}
  {\bf 30} 065009

\bibitem{Derzsi_2017}
Derzsi A, Bruneau B, Gibson A~R, Johnson E, O'Connell D, Gans T, Booth J~P and
  Donk{\'{o}} Z 2017 {\em Plasma Sources Science and Technology\/} {\bf 26}
  034002

\bibitem{Gudmundsson_2019}
Gudmundsson J~T and Proto A 2019 {\em Plasma Sources Science and Technology\/}
  {\bf 28} 045012

\bibitem{Donko_2017_PPCF}
Donk{\'{o}} Z, Derzsi A, Korolov I, Hartmann P, Brandt S, Schulze J, Berger B,
  Koepke M, Bruneau B, Johnson E, Lafleur T, Booth J~P, Gibson A~R, O'Connell D
  and Gans T 2018 {\em Plasma Physics and Controlled Fusion\/} {\bf 60} 014010

\bibitem{Brandt_2019}
Brandt S, Berger B, Donk{\'{o}} Z, Derzsi A, Schüngel E, Koepke M and Schulze
  J 2019 {\em Plasma Sources Science and Technology\/} {\bf 28} 095021

\bibitem{Hyo-Chang_2019_O2}
You K~H, Schulze J, Derzsi A, Donkó Z, Yeom H~J, Kim J~H, Seong D~J and Lee
  H~C 2019 {\em Physics of Plasmas\/} {\bf 26} 013503

\bibitem{Gibson_2017}
Gibson A~R and Gans T 2017 {\em Plasma Sources Science and Technology\/} {\bf
  26} 115007

\bibitem{Gans_2004}
Gans T, Schulz-von~der Gathen V and Döbele H~F 2004 {\em Contributions to
  Plasma Physics\/} {\bf 44} 523--528

\bibitem{Gans_2010}
Gans T, O'Connell D, von~der Gathen V~S and Waskoenig J 2010 {\em Plasma
  Sources Science and Technology\/} {\bf 19} 034010

\bibitem{Schulze_JPD_2010}
Schulze J, Schüngel E, Donk{\'{o}} Z, Luggenhölscher D and Czarnetzki U 2010
  {\em Journal of Physics D: Applied Physics\/} {\bf 43} 124016

\bibitem{Horvath_2020}
Horv{\'{a}}th B, Derzsi A, Schulze J, Korolov I, Hartmann P and Donk{\'{o}} Z
  2020 {\em Plasma Sources Science and Technology\/} {\bf 29} 055002

\bibitem{Hockney_Book}
Hockney R~W and Eastwood J~W 1988 {\em Computer Simulation Using Particles\/}
  (USA: Taylor \& Francis, Inc.) ISBN 0852743920

\bibitem{Birdsall_Book}
 C. K. Birdsall and A. B. Langdon, {\em Plasma Physics via Computer Simulation}
  (McGraw-Hill, New York, 1985).

\bibitem{Birdsall_1991}
{Birdsall} C~K 1991 {\em IEEE Transactions on Plasma Science\/} {\bf 19} 65--85
  ISSN 0093-3813

\bibitem{Schweigert_1999}
Schweigert V~A, Alexandrov A~L, Gimelshtein S~F and Ivanov M~S 1999 {\em Plasma
  Sources Science and Technology\/} {\bf 9} B1--B3

\bibitem{Diomede_2005}
Diomede P, Capitelli M and Longo S 2005 {\em Plasma Sources Science and
  Technology\/} {\bf 14} 459--466

\bibitem{Verboncoeur2005}
{Verboncoeur} J~P 2005 {\em Plasma Physics and Controlled Fusion\/} {\bf 47}
  A231--A260

\bibitem{Schneider}
Matyash K, Schneider R, Taccogna F, Hatayama A, Longo S, Capitelli M, Tskhakaya
  D and Bronold F 2007 {\em Contributions to Plasma Physics\/} {\bf 47}
  595--634

\bibitem{Radmilovic-Radjenovic2009}
Radmilovi{\'{c}}-Radjenovi{\'{c}} M and Petrovi{\'{c}} Z~L 2009 {\em The
  European Physical Journal D\/} {\bf 54} 445--449

\bibitem{Donko_2011_PSST}
Donk{\'{o}} Z 2011 {\em Plasma Sources Science and Technology\/} {\bf 20}
  024001

\bibitem{Donko2012}
Donk{\'{o}} Z, Schulze J, Czarnetzki U, Derzsi A, Hartmann P, Korolov I and
  Schüngel E 2012 {\em Plasma Physics and Controlled Fusion\/} {\bf 54} 124003

\bibitem{Gudmundsson_2013}
Gudmundsson J~T, Kawamura E and Lieberman M~A 2013 {\em Plasma Sources Science
  and Technology\/} {\bf 22} 035011

\bibitem{Sun_2016}
Sun A, Becker M~M and Loffhagen D 2016 {\em Computer Physics Communications\/}
  {\bf 206} 35 -- 44

\bibitem{Bronold_2007}
Bronold F~X, Matyash K, Tskhakaya D, Schneider R and Fehske H 2007 {\em Journal
  of Physics D: Applied Physics\/} {\bf 40} 6583--6592

\bibitem{Dittmann_2007}
Dittmann K, Drozdov D, Krames B and Meichsner J 2007 {\em Journal of Physics D:
  Applied Physics\/} {\bf 40} 6593--6600

\bibitem{Matyash_2007}
Matyash K, Schneider R, Dittmann K, Meichsner J, Bronold F~X and Tskhakaya D
  2007 {\em Journal of Physics D: Applied Physics\/} {\bf 40} 6601--6607

\bibitem{Bera_2011}
Bera K, Rauf S and Collins K 2011 {\em IEEE Transactions on Plasma Science\/}
  {\bf 39} 2576--2577

\bibitem{Teichmann_2013}
Teichmann T, Küllig C, Dittmann K, Matyash K, Schneider R and Meichsner J 2013
  {\em Physics of Plasmas\/} {\bf 20} 113509

\bibitem{Liu_2013}
Liu J, Wen D~Q, Liu Y~X, Gao F, Lu W~Q and Wang Y~N 2013 {\em Journal of Vacuum
  Science \& Technology A\/} {\bf 31} 061308

\bibitem{Liu_2015}
Liu J, Liu Y~X, Liu G~H, Gao F and Wang Y~N 2015 {\em Journal of Applied
  Physics\/} {\bf 117} 143301

\bibitem{Benyoucef_2015}
Benyoucef D and Yousfi M 2015 {\em Physics of Plasmas\/} {\bf 22} 013510

\bibitem{Kullig_2015}
Küllig C, Wegner T and Meichsner J 2015 {\em Physics of Plasmas\/} {\bf 22}
  043515

\bibitem{Hannesdottir_2016}
Hannesdottir H and Gudmundsson J~T 2016 {\em Plasma Sources Science and
  Technology\/} {\bf 25} 055002

\bibitem{Hannesdottir_2017}
Hannesdottir H and Gudmundsson J~T 2017 {\em Journal of Physics D: Applied
  Physics\/} {\bf 50} 175201

\bibitem{Gudmundsson_2017}
Gudmundsson J~T and Snorrason D~I 2017 {\em Journal of Applied Physics\/} {\bf
  122} 193302

\bibitem{Gudmundsson_2018}
Gudmundsson J~T, Snorrason D~I and Hannesdottir H 2018 {\em Plasma Sources
  Science and Technology\/} {\bf 27} 025009

\bibitem{Derzsi_2016}
Derzsi A, Lafleur T, Booth J~P, Korolov I and Donk{\'{o}} Z 2016 {\em Plasma
  Sources Science and Technology\/} {\bf 25} 015004

\bibitem{Wang_2020}
Wang L, Wen D~Q, Hartmann P, Donk{\'{o}} Z, Derzsi A, Wang X~F, Song Y~H, Wang
  Y~N and Schulze J 2020 {\em Plasma Sources Science and Technology\/} {\bf 29}
  105004

\bibitem{Vass_2020}
Vass M, Wilczek S, Lafleur T, Brinkmann R~P, Donk{\'{o}} Z and Schulze J 2020
  {\em Plasma Sources Science and Technology\/} {\bf 29} 025019

\bibitem{Derzsi_2015}
Derzsi A, Korolov I, Schüngel E, Donk{\'{o}} Z and Schulze J 2015 {\em Plasma
  Sources Science and Technology\/} {\bf 24} 034002

\bibitem{Donko_2018}
Donk{\'{o}} Z, Derzsi A, Vass M, Schulze J, Schuengel E and Hamaguchi S 2018
  {\em Plasma Sources Science and Technology\/} {\bf 27} 104008

\bibitem{Schulenberg21}
Schulenberg D~A, Korolov I, Donk\'o Z, Derzsi A and Schulze J 2021 {\em Plasma
  Sources Sci. Technol.\/} {\bf 30} 105003

\bibitem{T6389}
Kr{\"o}tz W, Ulrich A, Ribitzki G, Wieser J and Murnick D~E 1994 {\em Hyperfine
  Interactions\/} {\bf 88} 193--203

\bibitem{Saloman04}
Saloman E~B and Sansonetti C~J 2004 {\em Journal of Physical and Chemical
  Reference Data\/} {\bf 33} 1113--1158

\bibitem{Vass_2021}
Vass M, Wilczek S, Lafleur T, Brinkmann R~P, Donk{\'{o}} Z and Schulze J 2021
  {\em Plasma Sources Science and Technology\/} {\bf 30} 065015

\bibitem{Geonwoo_heating}
Park G, Kim J~S, Kim C~H, Kim H~J and Lee H~J 2022 {\em IEEE Transactions on
  Plasma Science\/} {\bf 50} 540--549

\bibitem{Biagi7.1}
 Biagi-v7.1 database, \url{www.lxcat.net}, retrieved on April 8, 2019. Cross
  sections extracted from PROGRAM MAGBOLTZ, VERSION 7.1 JUNE 2004.

\bibitem{Biagi8.9}
 Biagi-v8.9 database, \url{www.lxcat.net}, retrieved on November 25, 2014.
  Cross sections extracted from PROGRAM MAGBOLTZ, VERSION 8.9 March 2010.

\bibitem{vahedi1995monte}
Vahedi V and Surendra M 1995 {\em Computer Physics Communications\/} {\bf 87}
  179--198

\bibitem{gudmundsson2013benchmark}
Gudmundsson J~T, Kawamura E and Lieberman M~A 2013 {\em Plasma Sources Science
  and Technology\/} {\bf 22} 035011

\bibitem{PhepsNeonJILA}
Phelps A~V personal communication

\bibitem{adams1972thermal}
Adams N, Dean A and Smith D 1972 {\em International Journal of Mass
  Spectrometry and Ion Physics\/} {\bf 10} 63--76

\bibitem{Langevin}
 P. Langevin, {\em Une formule fondamentale de th\'eorie cin\'etique} (1905).

\bibitem{bronold2007radio}
Bronold F, Matyash K, Tskhakaya D, Schneider R and Fehske H 2007 {\em Journal
  of Physics D: Applied Physics\/} {\bf 40} 6583

\bibitem{schlumbohm1969dissoziativer}
Schlumbohm H 1969 {\em Zeitschrift f{\"u}r Naturforschung A\/} {\bf 24}
  1720--1724

\bibitem{Olney1997}
Olney T~N, Cann N, Cooper G and Brion C 1997 {\em Chemical Physics\/} {\bf 223}
  59--98 ISSN 0301-0104

\bibitem{pratt1979pulverized}
Pratt D~T, Smoot L and Pratt D 1979 {\em Pulverized coal combustion and
  gasification\/} (Springer)

\bibitem{matsumoto1991comparison}
Matsumoto H and Koura K 1991 {\em Physics of Fluids A: Fluid Dynamics\/} {\bf
  3} 3038--3045

\bibitem{khrapak2014accurate}
Khrapak S~A 2014 {\em The European Physical Journal D\/} {\bf 68} 1--6

\bibitem{schnabel2007unlike}
Schnabel T, Vrabec J and Hasse H 2007 {\em Journal of Molecular Liquids\/} {\bf
  135} 170--178

\bibitem{Magnusson2020}
Magnusson J~M, Collins A~L and Wirz R~E 2020 {\em Aerospace\/} {\bf 7}

\bibitem{Serikov97}
Serikov V~V and Nanbu K 1997 {\em Journal of Applied Physics\/} {\bf 82}
  5948--5957

\bibitem{Gombosi}
Gombosi T~I 1994 {\em Gaskinetic Theory\/} (Cambridge University Press) ISBN
  0521439663

\bibitem{Sazhin97}
Sazhin S and Serikov V 1997 {\em Planetary and Space Science\/} {\bf 45}
  361--368 ISSN 0032-0633

\bibitem{Kennard}
Kennard E~H 1938 {\em Kinetic Theory Of Gases With An Introduction To
  Statistical Mechanics\/} (New York, USA: McGraw-Hill)

\bibitem{Chapman1939}
Chapman S and Cowling T~G 1939 {\em Mathematical Theory of Non‐Uniform
  Gases\/} (London, England: Cambridge University Press)

\bibitem{Winn1950}
Winn E~B 1950 {\em Phys. Rev.\/} {\bf 80}(6) 1024--1027

\bibitem{StatPhys}
McQuarrie D~A 2000 {\em Statistical Mechanics\/} (Sausalito, California, USA:
  University Science Books) ISBN 1891389157

\bibitem{Schulze2018_Boltzmann}
{Schulze} J, {Donk{\'o}} Z, {Lafleur} T, {Wilczek} S and {Brinkmann} R~P 2018
  {\em Plasma Sources Science and Technology\/} {\bf 27} 055010

\bibitem{Schulze2014_ambipolar}
Schulze J, Donk{\'{o}} Z, Derzsi A, Korolov I and Schuengel E 2014 {\em Plasma
  Sources Science and Technology\/} {\bf 24} 015019

\bibitem{NumericPIC}
Vass M, Palla P and Hartmann P Revisiting the numerical stability/accuracy
  conditions of {PIC/MCC} simulations of low-temperature gas discharges,
  accepted for publication in Plasma Sources Science and Technology
  \urlprefix\url{https://doi.org/10.1088/1361-6595/ac6e85}

\end{thebibliography}

\providecommand{\newblock}{}

\end{document}